\documentclass{article}

\usepackage[titletoc, title]{appendix}

 \usepackage[top=1.in, bottom=1.in, left=0.9in, right=0.9in]{geometry}
\usepackage{ amssymb }
\usepackage{ mathtools }
\usepackage{graphicx} 
\usepackage{caption}
\usepackage{subcaption}
\usepackage{amsmath} 
\usepackage[title]{appendix}
\usepackage{authblk}
\usepackage{wrapfig}
\usepackage{natbib}
\setcitestyle{authoryear,open={(},close={)}}

\usepackage[font={small}]{caption}
\usepackage{amsmath}
\usepackage{tikz}

\usepackage{placeins}
\usepackage{hyperref}

\definecolor{deepcarmine}{rgb} {0.66, 0.13, 0.24}
\hypersetup{
	colorlinks=true,       
	linkcolor=black,          
	citecolor=deepcarmine,        
}

\title{Shaping dynamics with multiple populations in low-rank recurrent networks} 

\author[1]{Manuel \textsc{Beiran}}
\author[1]{Alexis \textsc{Dubreuil}}
\author[1]{Adrian \textsc{Valente}}
\author[2]{Francesca \textsc{Mastrogiuseppe}}
\author[1]{Srdjan \textsc{Ostojic}}

\affil[1]{Laboratoire de Neurosciences Cognitives et Computationnelles, INSERM U960, Ecole Normale Superieure - PSL University, 75005 Paris, France}
\affil[2]{Gatsby Computational Neuroscience Unit, UCL, London, Great Britain}

\date{\today} 

\begin{document}
\maketitle 
\begin{abstract}
    An emerging paradigm proposes that neural computations can be understood at the level of dynamical systems that govern low-dimensional trajectories of collective neural activity. How the connectivity structure of a network determines the emergent dynamical system however remains to be clarified. Here we consider a novel class of models, Gaussian-mixture low-rank recurrent networks, in which the rank of the connectivity matrix and the number of statistically-defined populations are independent hyper-parameters. We show that the resulting collective dynamics form a dynamical system, where the rank sets the dimensionality and the population structure shapes the dynamics. In particular, the collective dynamics can be described in terms of a simplified effective circuit of interacting latent variables. While having a single, global population strongly restricts the possible dynamics, we demonstrate that if the number of populations is large enough, a rank $R$ network can approximate any $R$-dimensional dynamical system.
\end{abstract}

\newpage

\section{Introduction}

A newly emerging paradigm posits that neural computations rely on collective dynamics in the state-space corresponding to the joint activity of all neurons in a network \citep{ Churchland2007, Rabinovich2008, Buonomano2009, Saxena2019, Vyas2020}. Experiments in behaving animals have found that trajectories of neural activity are typically restricted to low-dimensional manifolds in that space \citep{Machens2010, Mante2013, Rigotti2013, Gao2015, Gallego2018, Chaisangmongkon2017,  Wang2018, Sohn2019}, and can therefore be described by a small number of collective, latent variables. It has been proposed that these collective variables form  dynamical systems that implement computations through their responses to inputs \citep{Eliasmith2003, Hennequin2014, Rajan2016, Remington2018a, Remington2018b}. How synaptic connectivity shapes the effective dynamics of collective variables, and therefore computations, however remains to be clarified.

Recurrent neural networks (RNNs) trained to perform neuroscience tasks are an ideal model system to address this question and further develop the theory of computations through dynamics \citep{Sussillo2015, Rajan2016, Barak2017, Wang2018, Yang2019}. A recently introduced class of models, low-rank RNNs, directly embodies the idea of low-dimensional collective dynamics,  opens the door to relating connectivity and dynamics, and provides a framework that unifies a number of specific RNN classes \citep{Mastrogiuseppe2018}. Low-rank RNNs  rely on connectivity matrices that are restricted to be low rank, which directly generate low-dimensional  dynamics. The rank of the network determines the number of  collective variables needed to provide a full description of the collective dynamics. While previous works have shown that other specific classes of RNNs can approximate arbitrary dynamical systems \citep{Doya1993, Maass2007}, the range of collective dynamics that can be implemented by low-rank RNNs however remains to be clarified.

In this work, we focus on low-rank RNNs in which neurons are organized in distinct populations that correspond to clusters in the space of low-rank connectivity patterns. Each population is defined by its statistics of connectivity, described by a multi-variate Gaussian distribution, so that the full network is specified by a mixture of Gaussians.  The total number of populations in the network is a hyper-parameter distinct from the rank of connectivity. Previous works have considered low-rank networks consisting of a single, global Gaussian population \citep{Mastrogiuseppe2018, Mastrogiuseppe2019, Schuessler2020}. In the opposite limit, by increasing the number of populations, a Gaussian mixture model can approximate any arbitrary low-rank connectivity distribution. Here we examine how the number of populations and their structure determine and limit the resulting collective dynamics in the network.

We first derive three general properties of Gaussian-mixture low-rank networks: (i) in an autonomous network of rank $R$, dynamics are characterized by $R$ collective variables that form a dynamical system; (ii)  the dynamics are determined by an effective circuit description, where collective variables interact through gain-modulated effective couplings; (iii)  the resulting low-dimensional dynamics can approximate any arbitrary $R$-dimensional dynamical system if the number of populations is large enough. We then proceed to illustrate how increasing the number of populations in a network extends its dynamical range. For that, we specifically focus on fixed points of the dynamics. While a network consisting of a single population can generate at most a pair of stable fixed points, independently of its rank, we show that adding populations allow the network to implement arbitrary numbers of stable fixed points embedded in a subspace determined by the rank of the connectivity matrix.  Finally, we propose a general algorithm to approximate a given $R$-dimensional dynamical system with a multi-population  network of rank $R$, and show one example network that is designed to implement complex temporal dynamics.

\newpage

\section{Model class: Gaussian mixture low-rank networks}
    In this section, we introduce the class of models we study, and define the key underlying quantities.

     We consider a recurrent neural network of $N$ rate units. The dynamics of the input $x_i$ to the $i$-th unit are given by
    \begin{equation}\label{dyn1}
        \tau \frac{d x_i}{dt} = - x_i + \sum_{j=1}^N J_{ij} \phi\left(x_i\right) +  I_i^{ext}\left(t\right)
    \end{equation}
    \noindent where $\tau$ corresponds to the membrane time constant, the matrix element $J_{ij}$ is the synaptic strength from unit $j$ to unit $i$ and $I_i^{ext}\left(t\right)$ is the external input received by the $i$-th unit.  The non-linear function $\phi\left(x\right)$ maps the input of a neuron to its firing rate activity. Throughout this study, we use the non-linear activation function $\phi\left(x\right) = \tanh\left(x\right)$, although the theoretical results in Section~\ref{section:model} hold for any non-polynomial activation function. 
    
\begin{figure}[h]
     \centering
     \begin{subfigure}[b]{0.8\textwidth}
        \includegraphics[width=\linewidth]{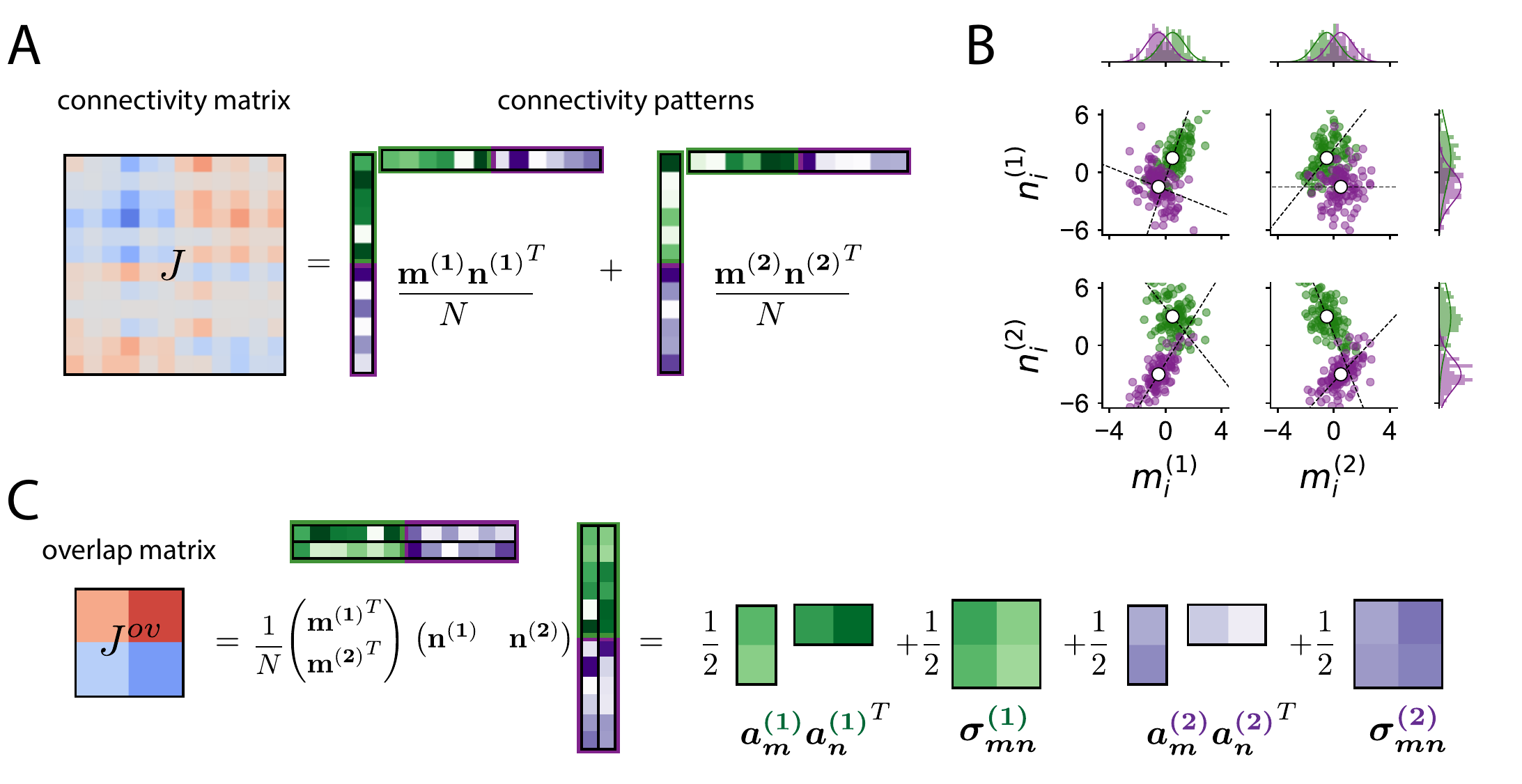}
     \end{subfigure}
	   	\caption{\textbf{Low-rank connectivity with Gaussian populations}. \textbf{A} The connectivity matrix $J$, rank-two in this illustration, is decomposed into the sum of two rank-one terms given by the outer product of the connectivity patterns $\mathbf{m^{\left(r\right)}}$ and $\mathbf{n^{\left(r\right)}}$, $r=1,2$. The components of the connectivity patterns -- the pattern loadings -- are grouped into two different sub-patterns (green and purple) with different population statistics. For visual purposes, the connectivity is shown only for 12 neurons in each population, the first 12 neurons belong to population 1 and the last 12 neurons belong to population 2. \textbf{B} Scatter plot of the distribution of pattern components in the four-dimensional loading space. Each dot corresponds to one neuron, and each neuron is characterized by its four values on the patterns $\mathbf{m^{\left(r\right)}}$ and $\mathbf{n^{\left(r\right)}}$, $r=1,2$. The color indicates whether the neuron belongs to the first population (green) or the second population (purple). The different populations are defined by different multivariate Gaussian statistics, means (white dots) and covariances (dashed lines), and define separate clusters. Population size $N=200$, $\alpha_p = 0.5$. \textbf{C} Overlap matrix given by the inner product between connectivity patterns. The overlap matrix is a square matrix of size given by the rank of the connectivity, in this case $2\times 2$. Its eigenvalues coincide with the non-zero eigenvalues of the $N \times N$ connectivity matrix. The overlap matrix can be expressed as a weighted sum over the overlaps of the different populations, as shown in Eq.~\eqref{ov_mat}. } \label{fig:T0}
    \end{figure}

    We restrict the connectivity matrix to be of low rank, i.e. the number of non-zero singular values of the matrix $J$ is $R\ll N$. Using singular value decomposition, any connectivity matrix of this type can be expressed as the sum of $R$ unit rank terms,
    \begin{equation}\label{connec}
        J_{ij} = \frac{1}{N} \sum_{r=1}^{R} m^{\left(r\right)}_i n^{\left(r\right)}_j.
    \end{equation}
    \noindent The connectivity is therefore characterized by a set of $R$  N-dimensional vectors, or connectivity patterns, $\mathbf{m^{(r)}}=\left\{m^{\left(r\right)}_i\right\}_{i=1\dots N}$ and $\mathbf{n^{(r)}}=\left\{n^{\left(r\right)}_i\right\}_{i=1\dots N}$ for $r=1,\dots, R$, where  $\mathbf{m^{(r)}}$
    are the left singular vectors of the connectivity matrix, and $\mathbf{n^{(r)}}$  correspond to the right singular vectors multiplied by the corresponding singular values (see Fig.~\ref{fig:T0} A for an example of a rank-two connectivity matrix). 
    The vectors $\mathbf{m^{(r)}}$ (resp. $\mathbf{n^{(r)}}$) for $r=1,\dots, R$ are mutually orthogonal.
    Without loss of generality, we fix the norm of the left singular vectors $\mathbf{m^{\left(r\right)}}$ to be equal to $N$. This decomposition is unique, up to a change in sign of the set of vectors $\mathbf{m^{\left(r\right)}}$ and $\mathbf{n^{\left(r\right)}}$. 
    
    The external input can be expressed as the sum of $N_{in}$ time-varying terms
    \begin{equation}
        I_i^{ext} \left(t\right) = \sum_{s=1}^{N_{in}} I_i^{\left(s\right)} u_s\left(t\right), 
    \end{equation}
    \noindent which are fed into the network through a set of orthonormal input patterns $\mathbf{I^{\left(s\right)}}=\left\{I_i^{\left(s\right)}\right\}_{i=1 \dots N}$ for $s = 1,\dots, N_{in}$. In this study, we focus on the dynamics of autonomous networks or networks with a constant external input.
   
    Each neuron in the network is therefore characterized by its $2R+N_{in}$ components on the connectivity patterns $\mathbf{m^{(r)}}$ and $\mathbf{n^{(r)}}$ and input patterns $\mathbf{I^{\left(s\right)}}$.
    By analogy with factor analysis, we refer to these components  as pattern loadings, and denote the set of loadings for neuron $i$ as 
    \begin{equation}
         \left(\left\{{m^{(r)}_i}\right\}_{r=1\ldots R}, \left\{{n^{(r)}_i}\right\}_{r=1\ldots R}, \left\{{I^{(s)}_i}\right\}_{s=1\ldots N_{in}} \right) \coloneqq \left(\underline{m}_i, \underline{n}_i, \underline{I}_i\right).
    \end{equation}
    
    \noindent Each neuron can thus be represented as a point in the loading space of dimension $2R + N_{in}$, and the connectivity of the full network can therefore be described as a set of $N$ points in this pattern loading space (see Fig.~\ref{fig:T0} B). 
    
    We assume that for each neuron, the set of pattern loadings is generated independently  from a multi-variate probability distribution $P\left(\underline{m}, \underline{n}, \underline{I}\right)$. We moreover restrict ourselves to a specific class of loading distributions,  mixtures of multi-variate Gaussians. This choice is motivated by the fact that Gaussian mixtures can approximate any arbitrary multi-variate distribution, afford a natural interpretation in terms of populations, and allow for a mathematically tractable and transparent analysis of the dynamics as shown below.
    
    In this Gaussian mixture model, each neuron is  assigned to a population $p$ with probability $\alpha_p$, $p=1\dots P$, so that the connectivity matrix $J$ is a block matrix. Within population $p$, the joint distribution $P^{\left(p\right)} \left(\underline{m}, \underline{n}, \underline{I} \right)$ is a multivariate Gaussian  defined by (i) its mean $\boldsymbol{a^{\left(p\right)}}$,  a vector of dimension $2R+N_{in}$, given by the set of means of each pattern loading within population $p$
     \begin{equation}\label{def_mu}
      \boldsymbol{a^{\left(p\right)}} = \left(a_{m_1}^{\left(p\right)}, \dots, a_{m_R}^{\left(p\right)}, a_{n_1}^{\left(p\right)}, \dots, a_{n_R}^{\left(p\right)}, a_{I_1}^{\left(p\right)}, \dots, a_{I_{N_{in}}}^{\left(p\right)}\right),   
     \end{equation}
     \noindent and (ii) its covariance  $\Sigma^{\left(p\right)}$, a matrix of dimension $(2R+N_{in}) \times (2R+N_{in})$, whose elements are  the pairwise covariances
     \begin{equation}\label{corr_gen}
         \Sigma_{xy}^{\left(p\right)} = E\left[\left(x^{\left(p\right)}-a_x^{\left(p\right)}\right)\left(y^{\left(p\right)}-a_y^{\left(p\right)}\right)\right]
     \end{equation}
     \noindent where $E\left[\,\cdot \,\right]$ indicates the expected value, and $x$ and $y$ represent any pair of connectivity or input components.  Within the loading space, each population therefore corresponds to a cluster centered at $\boldsymbol{a^{\left(p\right)}}$, and of shape specified by the connectivity matrix $\Sigma_{xy}^{\left(p\right)}$ (see Fig.~\ref{fig:T0} B).
    
    The geometrical arrangement between patterns is a key feature to understand the behavior of low-rank networks \citep{Mastrogiuseppe2018}. The connectivity and input patterns are $N$-dimensional vectors. To quantify the geometrical configuration between two patterns, we define the overlap, or normalized scalar product:
    \begin{equation}\label{ov1}
        O\left(\mathbf{x}, \mathbf{y}\right) = \frac{1}{N} \sum_{i=1}^N x_i y_i
    \end{equation}
    \noindent  where $\mathbf{x}$ and $\mathbf{y}$ are any two patterns in the set given by $\mathbf{m^{\left(r\right)}}, \mathbf{n^{\left(r\right)}}$ and $\mathbf{I^{\left(s\right)}}$. The overlap is the projection of pattern $\mathbf{x}$ onto $\mathbf{y}$, so that two patterns are orthogonal if and only if their overlap is zero.
    
      An important property of rank-$R$ matrices, such as the connectivity matrix $J$, is that their non-zero eigenvalues coincide with the eigenvalues of the overlap matrix $J^{ov}$ \citep{Nakatsukasa2019} that is defined by the overlaps between pairs of connectivity patterns: 
    \begin{equation}\label{def_eig}
        J_{rs}^{ov} = O\left(\mathbf{m^{\left(s\right)}}, \mathbf{n^{\left(r\right)}}\right),
    \end{equation}
\noindent for $r, s = 1, \dots, R$. The eigenvalues of the connectivity matrix, and therefore of the overlap matrix, are an essential property to understand the dynamics of low-rank networks, as we show in Section 4. It is often more convenient to calculate the eigenspectrum of the overlap matrix $J^{ov}$, of size $R\times R$, than of the connectivity matrix $J$, of size $N\times N$.

    In a network with $P$ populations, any pattern $\mathbf{x}$ of length $N$ can be represented as a set of $P$ sub-patterns $\mathbf{x^{\left(p\right)}}$, for $p=1,\dots,P$, where each sub-pattern has length $\alpha_p N$ and includes the components of neurons belonging to population $p$. Fig.~\ref{fig:T0} shows an example of a rank-two network with two populations, where the connectivity patterns can be split into two different sub-patterns of equal size (green and purple). The overlap between two patterns can then be expressed as a weighted average of the overlaps between sub-patterns:
    \begin{equation}\label{ov2}
        O\left(\mathbf{x}, \mathbf{y}\right) = \sum_{p=1}^P \alpha_p O\left(\mathbf{x^{\left(p\right)}}, \mathbf{y^{\left(p\right)}}\right).
    \end{equation}    
    \noindent Even if the sub-patterns are not orthogonal to each other, i.e. the overlap between two sub-patterns is not zero, the patterns can be orthogonal to each other when the sub-pattern overlaps cancel out.                         
    In the limit of large networks, the overlap between two sub-patterns $\mathbf{x^{\left(p\right)}}$ and $\mathbf{y^{\left(p\right)}}$ is given by the expected value over the distribution of the loadings in the population:
    \begin{equation}\label{overlap}
        O\left(\mathbf{x^{\left(p\right)}}, \mathbf{y^{\left(p\right)}}\right) = E\left[x^{\left(p\right)} y^{\left(p\right)}\right] = a_x^{\left(p\right)} a_y^{\left(p\right)} + \Sigma_{xy}^{\left(p\right)}.
    \end{equation}

    In order to define the overlap matrix in terms of the statistics of the different Gaussian populations, we define the matrix 
    \begin{equation}\label{s_mn}
     \sigma_{n_r m_s}^{\left(p\right)} = \Sigma_{m_s n_r}^{\left(p\right)}.
    \end{equation}
    \noindent  The matrix $\boldsymbol{\sigma_{mn}}^{\left(p\right)}$ is a $R\times R$ whose entries contain the covariance between the connectivity patterns $\mathbf{m^{\left(r\right)}}$ and the $\mathbf{n^{\left(r\right)}}$ in population $p$. We call this matrix $\boldsymbol{\sigma_{mn}}^{\left(p\right)}$ a (reduced) covariance matrix, in an abuse of notation, because it is a subset of the covariance matrix $\Sigma^{\left(p\right)}$, and therefore it is not symmetric nor positive definite. For example, for a rank-one network, $\boldsymbol{\sigma_{mn}}^{\left(p\right)}$ is just a scalar, that can take any real value. For a rank-two network, $\boldsymbol{\sigma_{mn}}^{\left(p\right)}$ is a $2\times 2$ matrix, whose entries are given by the four covariances $\sigma_{m_1 n_1}^{\left(p\right)}$, $\sigma_{m_1 n_2}^{\left(p\right)}$,  $\sigma_{m_2 n_1}^{\left(p\right)}$, and  $\sigma_{m_2 n_2}^{\left(p\right)}$. 

    Using Eqs.~\eqref{ov2} and \eqref{overlap}, we can characterize the overlap matrix $J^{ov}$ as a function of the statistics of the connectivity sub-patterns:
    \begin{equation}\label{ov_mat}
        J^{ov} = \sum_{p=1}^P \alpha_p \left( \boldsymbol{a_{n}^{\left(p\right)}} {\boldsymbol{a_{m}^{\left(p\right)}} }^T + \boldsymbol{\sigma_{mn}^{\left(p\right)}}\right),
    \end{equation}
    \noindent where $\boldsymbol{a_{n}^{\left(p\right)}}$ and $\boldsymbol{a_{m}^{\left(p\right)}}$ are $R$ dimensional vectors whose entries correspond to the corresponding subset of elements in $\boldsymbol{a^{\left(p\right)}}$ (Fig.~\ref{fig:T0} C).
    
    Similarly to the covariance matrix $\boldsymbol{\sigma_{mn}}$ that measures the correlations between connectivity patterns $\mathbf{m^{\left(r\right)}}$ and $\mathbf{n^{\left(r\right)}}$, we define the covariance $\mathbf{\sigma_{nI}}$ between the connectivity patterns $\mathbf{n^{\left(r\right)}}$ and the constant external input $\mathbf{I}$, as a vector of length $R$, where each component is defined as
    \begin{equation}
        \sigma_{n_r I}^{\left(p\right)} = \Sigma^{\left(p\right)}_{n_r I}
    \end{equation}
    \noindent for $r=1,\dots, R$. We assume that the input loadings and loadings of the left connectivity patterns are uncorrelated within each pattern, $\sigma_{m_r I}^{\left(p\right)}=0$.
    
    \FloatBarrier
    \section{Dynamics in Gaussian mixture low-rank networks}
    \label{section:model}
    
    In this section, we present three key properties of dynamics in mixture of Gaussian low-rank networks: (i) in a network of rank $R$, dynamics can be characterized by $R$ collective variables that form a dynamical system; (ii) for loadings drawn from Gaussian mixture distributions, the dynamics can be further described as an effective circuit in which collective variables interact through gain-modulated effective couplings; (iii) with a sufficient number of populations, the resulting low-dimensional dynamics can approximate an arbitrary $R$-dimensional dynamical system.
    
    Details of the derivations are provided in appendices \ref{app:1} and \ref{app:2}.
    
    \subsection{Low-dimensional dynamics}
    
     In recurrent networks with low-rank connectivity, the dynamics of the trajectories $\mathbf{x}\left(t\right)$ are embedded in a linear subspace of dimension $R+N_{in}$ spanned by the left singular vectors $\mathbf{m^{\left(r\right)}}$ and the external input patterns $\bf{I^{\left(s\right)}}$, and can therefore be expressed as
     
     \begin{equation}\label{nbasis}
    x_i\left(t\right) = \sum_{r=1}^R \kappa_r m_i^{\left(r\right)} + \sum_{s=1}^{N_{in}} \kappa_{I_s} I_i^{\left(s\right)}.
\end{equation}
     Here $\kappa_r$ and $\kappa_{I_s}$ are collective variables that are obtained by projecting the activity $\mathbf{x}\left(t\right)$ on the patterns $\bf{m^{\left(r\right)}}$ and $\bf{I^{\left(s\right)}}$, that we assume orthogonal to each other. Introducing the trajectory $\mathbf{x}\left(t\right)$ expressed in this new basis into Eq.~\eqref{dyn1}, the dynamics of the collective variables are then given by the following dynamical system:
    \begin{align}\label{dyn_kap_nopop1}
        \tau \frac{d\kappa_r}{dt} &= -\kappa_r + \kappa_r^{rec}\\
        \tau \frac{d\kappa_{I_s}}{dt} &= -\kappa_{I_s} + u_s\left(t\right) \nonumber\\
        \kappa_r^{rec} &= \frac{1}{N}\sum_{i=1}^N  n_i^{\left(r\right)} \phi\left(\sum_{s=1}^{N_{in}} I_i^{\left(s\right)}\kappa_{I_s}  + \sum_{l=1}^R m_i^{\left(l\right)} \kappa_l\right). \label{rec_dyn}
    \end{align}
    
    We focus in the following on networks receiving a  constant input, so that there is only one collective variable $\kappa_I$ along the input dimension, the value of which is constant. The recurrent connectivity contributes to the dynamics of $\kappa_r$  through the term $\kappa_r^{rec}$.

The dynamics of collective variables in Eq.~\eqref{dyn_kap_nopop1} are valid for any finite-size low-rank network, without any assumption on the values of pattern loadings. We next turn to networks where the pattern loadings are generated from specific distributions.

    \subsection{Dynamics in multi-population networks}
    For low-rank networks in which pattern loadings are generated for each neuron from a Gaussian mixture distribution, in the limit of large $N$ the dynamics in Eq.~\eqref{dyn_kap_nopop1} can be expressed in terms of the statistics of pattern loadings over the populations, and become (see appendix \ref{app:1}):
    
        \begin{align} \label{dyn_pop_general_cont}
        \tau \frac{d \kappa_r}{d t} = &- \kappa_r + \kappa_r^{rec}\\
        \kappa_r^{rec} = &\sum_{p=1}^P \alpha_p \left[a_{n_r}^{\left(p\right)}\left\langle \phi \left(\mu^{\left(p\right)}, \Delta^{\left(p\right)} \right)\right\rangle  + \left(\sigma_{n_r I}^{\left(p\right)} \kappa_I +\sum_{s=1}^R \sigma_{n_r m_s }^{\left(p\right)} \kappa_s\right) \left\langle \phi^\prime \left(\mu^{\left(p\right)}, \Delta^{\left(p\right)} \right) \right\rangle \right]. \label{rec_dyn_pop}
    \end{align}
    \noindent Here $\mu^{\left(p\right)}$ and  $\Delta^{\left(p\right)}$ are the mean and variance of  input to population $p$, given by
    \begin{eqnarray}\label{eq:mu-delta}
    \mu^{\left(p\right)} &=& a_I^{\left(p\right)} \kappa_I + \sum_{s=1}^R a_{m_s}^{\left(p\right)} \kappa_s\\
    \Delta^{\left(p\right)} &=& \sigma_{I^2}^{\left(p\right)} \kappa_I^2 +\sum_{r=1}^R  \sigma_{m_{r}^2}^{\left(p\right)} \kappa_{r}^2.
    \end{eqnarray}
    In  Eq.~\ref{rec_dyn_pop}, we used the Gaussian integral notation:
    \begin{equation}\label{notation}
    \left\langle f \left(\mu, \Delta\right)\right\rangle = \int dx\, \left(2\pi\right)^{-\frac{1}{2}} e^{-x^2/2} f\left(\mu + \sqrt{\Delta} x\right).
    \end{equation}
    
    \noindent The Gaussian integral notation $\left\langle f \left(\mu, \Delta\right)\right\rangle$ represents the expected value of the random variable obtained after applying the function $f$ to a random Gaussian variable characterized by mean $\mu$ and variance $\Delta$.

    The factor $\left\langle \phi^\prime \left(\mu^{\left(p\right)}, \Delta^{\left(p\right)}\right)\right\rangle$ in Eq.~\eqref{rec_dyn_pop} corresponds to the average gain of neurons in population $p$ in a given state, specified by the mean $\mu^{\left(p\right)}$ and variance $ \Delta^{\left(p\right)}$ of the inputs to the population $p$. For each population, this average gain multiplies the covariances $\sigma_{m_l n_r}^{\left(p\right)}$ and $\sigma_{n_r I}^{\left(p\right)}$,
    and the corresponding average over populations defines
    an effective connectivity 
    \begin{equation}\label{eff_coup}
        \Tilde{\sigma}_{x y} = \sum_{p=1}^{P} \alpha_p \sigma_{x y}^{\left(p\right)} \left\langle \phi^\prime \left(\mu^{\left(p\right)}, \Delta^{\left(p\right)}\right)\right\rangle.
    \end{equation}
    The contributions of the first-order statistics $a_{n_r}^{\left(p\right)}$ to the recurrent dynamics   are modulated by the average firing rate in population $p$, and define an effective input
    \begin{equation}
        \Tilde{a}_{n_r} = \sum_{p=1}^{P} \alpha_p a_n^{\left(p\right)} \left\langle \phi \left(\mu^{\left(p\right)}, \Delta^{\left(p\right)}\right)\right\rangle.
    \end{equation}
    
    \noindent Introducing the effective connectivity and inputs into Eq.~\eqref{dyn_pop_general_cont}, the dynamics of a low-rank network with uncorrelated constant input take the simple form of an effective circuit of interacting collective variables:
    \begin{equation}\label{eq:circuit}
        \tau \frac{d\kappa_r}{dt} = -\kappa_r + \Tilde{a}_{n_r}+\sum_{l=1}^R \Tilde{\sigma}_{ n_r m_l} \kappa_l.
    \end{equation}
    \noindent Note that Eq.~\eqref{eq:circuit} describes the full non-linear dynamics in the limit $N \to \infty$. Although the collective variables interact linearly through the effective connectivity and inputs, those depend implicitly on $\kappa_r$. The overall dynamics are therefore non-linear, the non-linearity being fully encapsulated in the effective inputs and couplings.

    \subsection{Universal approximation of low-dimensional dynamical systems}
    
    By mapping the dynamics in Eqs.~\eqref{dyn_pop_general_cont} and \eqref{eq:circuit} to a feed-forward network with a single hidden layer, and exploiting the universal approximation theorem \citep{Cybenko1989, Leshno1993}, we can show that
    a Gaussian mixture network of rank $R$  receiving a constant input  is a universal approximator of $R$-dimensional dynamical systems (Appendix~\ref{app:2}). More precisely, for a sufficient number of populations, the low-rank dynamics in Eq.~\eqref{rec_dyn_pop} and \eqref{eq:circuit} can approximate with arbitrary precision any $R$-dimensional dynamical system 
    \begin{equation}
        \frac{d\boldsymbol{\kappa}}{dt} = G\left(\boldsymbol{\kappa}\right),
    \end{equation}
     defined by a vector field
    
    \begin{equation}\label{flow}
        G\left(\left\{ \kappa_r \right\}_{r=1\dots R}\right):= \left(G_1\left(\left\{\kappa_r\right\}_{r=1\dots R}\right), \dots, G_R\left(\left\{\kappa_r\right\}_{r=1\dots R}\right)\right)
    \end{equation}
    
    \noindent over an arbitrary finite domain $\left\{\kappa_r\right\}_{r=1\dots R} \in \left[\kappa_r^{min}, \kappa_r^{\max}\right]$. More specifically, this result requires that  the vector field $G$ is bounded and piecewise continuous, and the transfer function is not a polynomial (Appendix~\ref{app:2}). 
    
    As an alternative to approximating any vector field over a bounded domain, we show that if the transfer function is bounded and monotonic, a rank-R network with multiple populations can approximate any vector field $G\left(\left\{\kappa_r\right\}_{r=1\dots R}\right)$ over the full domain of the collective variables, $\left\{\kappa_r\right\}_{r=1\dots R} \in \left[-\infty,+\infty\right]$, with the restriction that the vector field follows asymptotic leaky dynamics for large input values:
    
    \begin{equation}\label{leaky}
        \lim_{\kappa_s \to \pm \infty} \frac{\partial G_r}{\partial \kappa_{r^\prime}} \left(\kappa_1,...,\kappa_r\right) = -\delta_{r r^\prime}
    \end{equation}
    \noindent for any values $s, r, r^\prime =1, \dots, R$, where $G_r$ represents the $r$-th component of the vector field as in Eq.~\eqref{flow}, and $\delta_{ij}$ is the Kronecker delta. This stems from the fact that for large values of $\kappa_r$, the recurrent dynamics (Eq.~\ref{rec_dyn_pop}) saturate to a constant value.

    Note that the universal approximation theorem does not state how many populations $P$ are required to implement a given dynamical system, and does not provide an algorithm for finding the statistics of the different populations. 
    

\section{Dynamics in networks with a single population}
Having shown that a rank $R$ network with an arbitrary number of populations can approximate any $R$-dimensional dynamical system, we now illustrate how having a small number of populations in contrast limits the possible dynamics.

 We focus first on the case of networks consisting of a single Gaussian population. For simplicity, we focus on  autonomous networks, with zero-mean connectivity patterns. Specifically, we show that, independently of their rank, the range of dynamics such networks can implement is restricted.  This case was previously studied for connectivities that combined a rank-one or rank-two structure and a
 full-rank random component  \citep{Mastrogiuseppe2018, Schuessler2020}. Here we provide an overview of those results, and extend them to single-population networks of arbitrary rank. The fact that we focus on networks whose connectivity is low-rank allows us to provide a deeper analysis of the dynamics.
    
    In vectorial form, assuming zero-mean connectivity patterns, the collective dynamics in Eq.~\eqref{dyn_pop_general_cont} for one population read
    \begin{equation}\label{dyn_vec}
        \tau \frac{d\boldsymbol{\kappa}}{dt} = - \boldsymbol{\kappa} + \left\langle \phi^\prime \left(0, \boldsymbol{\kappa}^T \boldsymbol{\kappa}\right) \right\rangle\boldsymbol{\sigma_{mn}} \boldsymbol{\kappa} ,
    \end{equation}
    \noindent where we used the vector of collective variables $\boldsymbol{\kappa} \in \mathcal{R}^R$, and the $R \times R$ covariance matrix $\boldsymbol{\sigma_{mn}}$ as defined in Eq.~\eqref{s_mn}, which is equal to the overlap matrix (Eq.~\ref{ov_mat}) in the case of zero-mean connectivity patterns. In the following analysis, we show that the eigenvalues of the covariance matrix $\boldsymbol{\sigma_{mn}}$, which for $N \to \infty$ are identical to the eigenvalues of the connectivity matrix,  determine the dynamics in collective space. \cite{Schuessler2020} performed a similar analysis for networks with random connectivity and rank-one and rank-two perturbations. 
     
\paragraph{Fixed points}
     The fixed points of Eq.~\ref{dyn_vec} are given by
    \begin{equation} \label{dyn_vec_fp}
        \boldsymbol{\kappa}_0 = \left\langle \phi^\prime \left(0, \boldsymbol{\kappa}_0^T \boldsymbol{\kappa}_0\right) \right\rangle\boldsymbol{\sigma_{mn}} \boldsymbol{\kappa}_0. 
    \end{equation}
    
     For $\phi(x)=\tanh\left(x\right)$, the trivial point $\boldsymbol{\kappa_0}=0$ is always a solution. There might  however be non-trivial fixed points depending on the eigenvalues of the covariance matrix $\boldsymbol{\sigma_{mn}}$. The covariance matrix can have up to $R$ eigenvalues, that we denote $\lambda_r$, with associated eigenvectors $\boldsymbol{u_r}$. Each real and non-degenerate eigenvalue $\lambda_r$ of the covariance $\boldsymbol{\sigma_{mn}}$ generates a fixed point $\boldsymbol{\kappa_0^{\left(r\right)}} = \rho_r \boldsymbol{u_r}$, where $\rho_r$ is the radial distance of the fixed point along the direction set by the eigenvector $\boldsymbol{u_r}$. Introducing this parametrization of the fixed points in Eq.~\ref{dyn_vec_fp}, we obtain the following implicit equation for the value $\rho_r$:

    \begin{equation}\label{scal_fp}
        1 = \lambda_r \left\langle \phi^\prime \left(0,  \rho_r^2\right)\right\rangle .
    \end{equation}
    
\begin{figure}[h]
     \centering
     \begin{subfigure}[b]{\textwidth}
        \includegraphics[width=\linewidth]{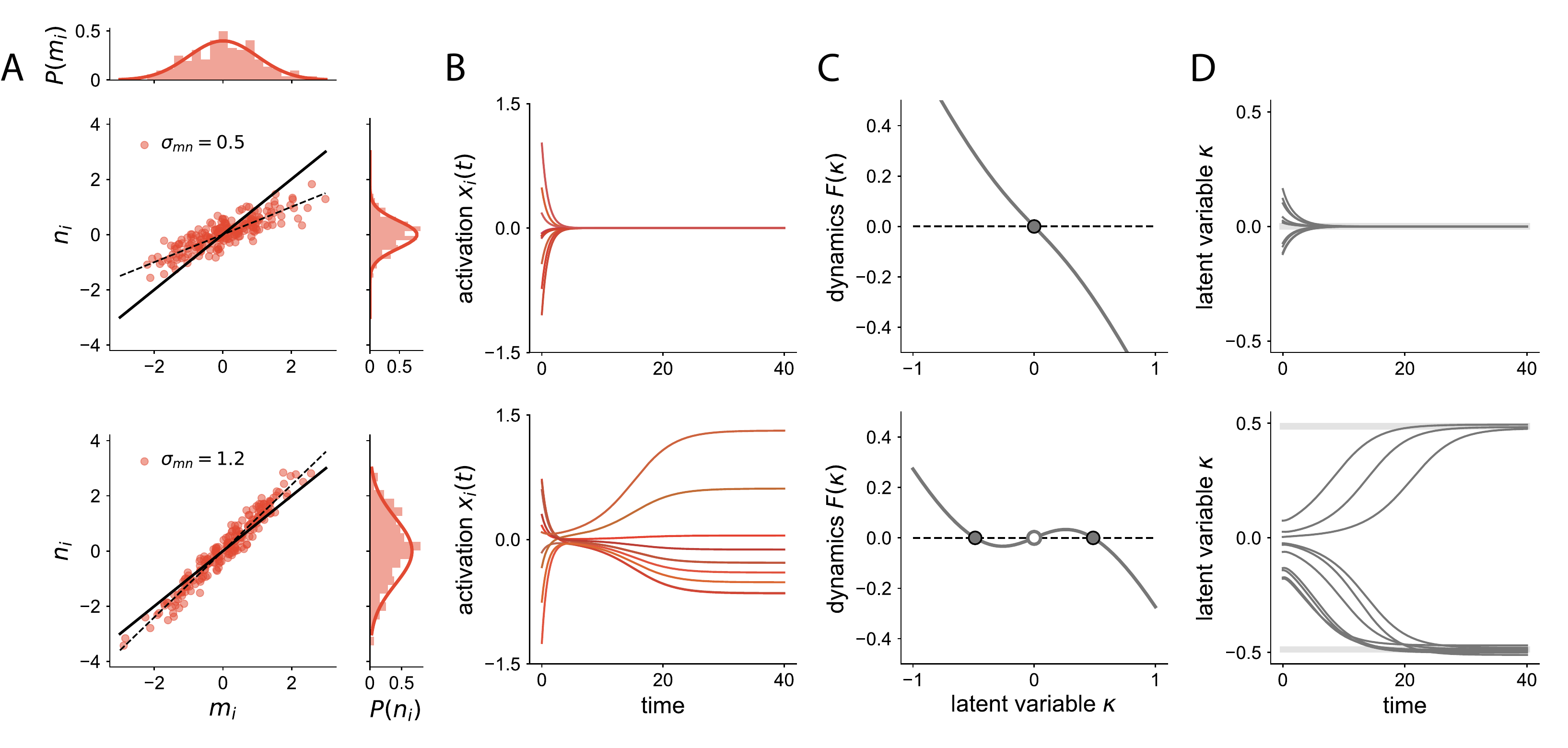}
     \end{subfigure}
	   	\caption{\textbf{Dynamics in rank-one networks with a single Gaussian population}. \textbf{A} Scatter plot of the loadings of left singular vectors $m_i^{\left(r\right)}$ and right singular vectors $n_i^{\left(r\right)}$. Top: Covariance $\sigma_{mn}$, indicated by the slope of the dashed line, below the critical value for non-trivial fixed points (solid line). Bottom: Covariance $\sigma_{mn}$ beyond the critical value.  \textbf{B} Dynamics of the activation variable $x_i\left(t\right)$ of ten units in the network for the two different networks initialized at random values. The network with $\sigma_{mn}$ larger than 1 (bottom) converges to a heterogeneous fixed point, while the other one decays to zero.  \textbf{C} One dimensional dynamics corresponding to the right hand side of Eq.~\eqref{dyn_vec}. Filled dots correspond to stable fixed points. For a weak covariance between connectivity patterns (top), the trivial fixed point is the only fixed point. For a strong covariance (bottom), the recurrent connectivity generates two non-trivial stable fixed points. \textbf{D} Evolution of the collective variable $\kappa$ as a function of time in a finite-size network, defined as the projection of the activity $\mathbf{x}\left(t\right)$ onto the connectivity pattern $\mathbf{m}$. Each curve corresponds to a different realization of the random connectivity matrix. $N=1000$, top row: $\sigma_{n^2} = 0.34$, bottom row $\sigma_{n^2} = 1.52$.} \label{fig:T1}
    \end{figure}

     \noindent The gain factor $\left\langle \phi^\prime \left(0, \rho_r^2\right)\right\rangle$ is bounded between 0 and 1 for the transfer function $\phi\left(x\right)=\tanh x$. Therefore, eigenvalues $\lambda_r>1$ generate two non-trivial fixed points, symmetrically located around the origin (see Fig.~\ref{fig:T1} A-D, bottom row, for a rank-one example). Smaller eigenvalues do not generate any non-trivial fixed point (Fig.~\ref{fig:T1} A-D, first row). 

     In order to determine the stability of the fixed points, we linearize the dynamics and obtain the Jacobian $S_r$  at the fixed point corresponding to the eigenvalue $\lambda_r$ of $\boldsymbol{\sigma_{mn}}$ (see appendix \ref{app:3})
     
     \begin{equation}\label{lin_stab}
         S_r = -\boldsymbol{I} +   \frac{1}{\lambda_r} \boldsymbol{\sigma_{mn}} + \left\langle \phi^{\prime\prime\prime} \left(0, \rho_r^2\right) \right\rangle \lambda_r \rho_r^2 \boldsymbol{u_r}\boldsymbol{u_r}^T,
     \end{equation}
     \noindent where $\boldsymbol{I}$ denotes the $R \times R$ identity matrix. The eigenvalues of $S_r$ determine the stability of the fixed points: if any positive eigenvalue exists, the dynamics will diverge away from the fixed point in the direction of the corresponding eigenvector of $S_r$. Negative eigenvalues correspond to attractive modes of the dynamics around the fixed point. If all eigenvalues of the stability matrix are negative, the fixed point is  stable. 
     
\begin{figure}[h]
     \centering
     \begin{subfigure}[b]{\textwidth}
        \includegraphics[width=\linewidth]{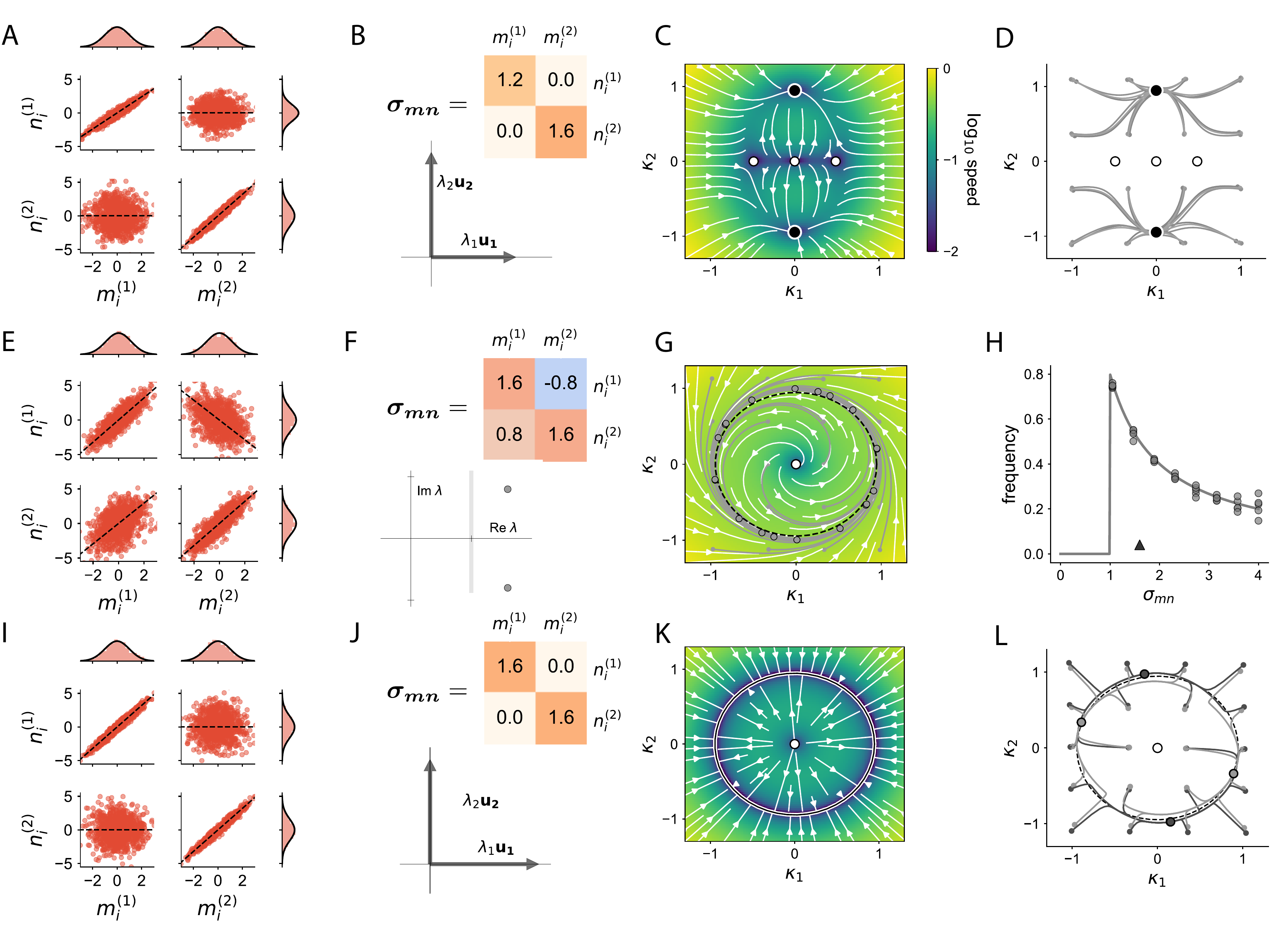}
     \end{subfigure}
     	   	\caption{\textbf{Dynamics in rank-two networks with a single Gaussian population - Connectivity matrix with real eigenvalues}. \textbf{A} Scatter plot of the loadings of left singular vectors $m_i^{\left(r\right)}$ and right singular vectors $n_i^{\left(r\right)}$. \textbf{B} Covariance matrix $\boldsymbol{\sigma_{mn}}$ of the population (top), and its eigenvectors (bottom).  \textbf{C} Vector field corresponding to the mean-field dynamics in the plane $\kappa_1-\kappa_2$ of collective variables (Eq.~\ref{dyn_vec}). The colormap represents the speed of the dynamics, defined as the norm of vector $\frac{d\boldsymbol{\kappa}}{dt}$, in different points of the collective space. Two non-trivial fixed points are generated in the direction of each eigenvector. Black dots correspond to stable fixed points, while white dots are unstable or saddle points. The pair of fixed points corresponding to the largest eigenvalue is stable. \textbf{D} Finite-size simulations of the dynamics. Three different connectivity realizations are shown from each initial condition. $N=1000$. \textbf{E-G} Similar to \textbf{A-C} for a network with complex eigenvalues (overlap matrix given by Eq.~\eqref{comp_corr}). The network generates a limit cycle. Grey curves in \textbf{G} correspond to trajectories from finite-size networks. The dots represent the final state after a fixed time elapsed. \textbf{H} Frequency of the limit cycle for different values of the symmetric part of the connectivity $\sigma$ and fixed imaginary part $\sigma_\omega = 0.8$. The dots show the numerically estimated frequency of oscillations in finite-size simulations for five different network realizations. The line corresponds to Eq.~\eqref{freq_LC}. The triangle indicates the parameter $\sigma$ used in \textbf{E-G}. \textbf{I-L} Similar to \textbf{A-D} for a network with degenerate eigenvalues: any vector in the plane spanned by vectors $\boldsymbol{m}^{\left(1\right)}$ and $\boldsymbol{m}^{\left(2\right)}$ is an eigenvector of the connectivity. This symmetry leads to a continuous attractor in the mean-field dynamics. In finite size simulations (two connectivity matrix realizations shown in \textbf{L}, in different shades of grey. Filled points correspond to the stable fixed points) the continuous attractor corresponds to a slow  manifold on which usually two stable fixed points lie. Parameters: $\sigma_{n_1^2} = 1.24$,  $\sigma_{n_2^2} = 1.63$.}
	   	 \label{fig:T2}
    \end{figure}

     When the eigenvectors of the matrix $\boldsymbol{\sigma_{mn}}$ are orthogonal to each other (Fig.~\ref{fig:T2} A-D), the $R$ eigenvalues of the matrix $S_r$, denoted as $\gamma_{r^\prime}$ for $r^\prime = 1 \dots R$, can  be calculated analytically: the eigenvalue $\gamma_{r^\prime}$ has an associated eigenvector equal to the eigenvector $\boldsymbol{u_{r^\prime}}$ of the covariance matrix $\boldsymbol{\sigma_{mn}}$, and reads 
     
     \begin{equation}\label{stab}
         \gamma_{r^\prime} = -1 + \frac{\lambda_{r^\prime}}{\lambda_r} + \left\langle \phi^{\prime\prime\prime} \left(0, \rho_r^2\right) \right\rangle \lambda_r \rho_r^2 \delta_{r r^\prime}.
     \end{equation}
     
    \noindent Remarkably, the eigenvalues of the Jacobian around any non-trivial fixed point are therefore directly determined by the eigenvalues of connectivity and covariance matrices \citep{Schuessler2020}.
    If $r^\prime=r$, the two first terms cancel out, and the third term is always negative (see appendix \ref{app:3}). This implies that all non-trivial fixed points are stable in the direction $\boldsymbol{u_r}$ that points  towards the origin. However, if there are other non-trivial fixed points  corresponding to eigenvalues $\lambda_{r^\prime}>\lambda_{r}$ of $\boldsymbol{\sigma_{mn}}$, the fixed point $\boldsymbol{\kappa_0^{\left(r\right)}}$ is destabilized in the directions of the eigenvectors with larger eigenvalues. 
    
    When the eigenvectors of $\boldsymbol{\sigma_{mn}}$ are not orthogonal to each other, the eigenvectors of $\boldsymbol{\sigma_{mn}}$ are not eigenvectors of the linear stability matrix $S_r$. However, the eigenvalues of $S_r$ are still given by Eq.~\eqref{stab} (see Appendix ~\ref{app:4}), so that the same stability properties hold (see Appendix D): every fixed point is stable in the direction towards the origin, and the fixed point in the direction given by the largest eigenvalue is stable, while the other ones become unstable.
     
     In summary, if all eigenvalues of the covariance matrix are real and non-degenerate, only the pair of non-trivial fixed points corresponding to the largest eigenvalue is stable. All the other non-trivial fixed points of the dynamics are saddle points. This implies that low-rank networks consisting of a single Gaussian population can have at most two stable fixed points independently of their rank.
    
\paragraph{Limit cycles}
     Complex eigenvalues of the covariance matrix $\boldsymbol{\sigma_{mn}}$, if they exist, always appear in conjugate pairs. They lead to spiral dynamics around the origin, in the plane spanned by the real and imaginary part of the corresponding eigenvectors. If the real part of the complex eigenvalues is smaller than unity, $\text{Re}\left(\lambda_r\right)<1$, the spiral dynamics decay back to the origin. Otherwise, if $\text{Re}\left(\lambda_r\right)>1$, there is a limit cycle on the plane, around the origin. Similarly to the case with only real eigenvalues of the covariance matrix, if the real part of the complex eigenvalue is larger than the real part of any other eigenvalue of $\boldsymbol{\sigma_{mn}}$, any trajectory will converge to the plane defined by the real and imaginary parts of the corresponding eigenvectors. On this plane, we then find that the limit cycle is stable. 
     
     To illustrate this case, we consider a rank two network with a covariance matrix of the form 
     \begin{equation}\label{comp_corr}
\boldsymbol{\sigma_{mn}} = 
\begin{pmatrix}
\sigma & -\sigma_\omega  \\
\sigma_\omega & \sigma
\end{pmatrix},
\end{equation}
    \noindent which has eigenvalues $\sigma \pm i \sigma_\omega$. Fig.~\ref{fig:T2} E-F shows an example of a network with such connectivity. 
    
    We can then write the equations for a rank-two network in polar form. Any state $\boldsymbol{\kappa} = \left[\kappa_1, \kappa_2\right]$ in a rank-two network  can be mapped to the radial distance $\rho$ and an orientation $\theta$ using the mapping $\kappa_1 := \rho \cos \theta$ and $\kappa_2 := \rho \sin \theta$. The dynamics in Eq.~\eqref{dyn_vec} become
    \begin{align}\label{rho}
        \tau \frac{d\rho}{dt} &= -\rho + \rho \sigma \left\langle \phi^\prime\left(0, \rho^2\right)\right\rangle\\ 
        \tau \frac{d\theta}{dt} &= \sigma_{\omega} \left\langle \phi^\prime\left(0, \rho^2\right)\right\rangle. \label{ang}
    \end{align}
     
     When the real part $\sigma$ of the eigenvalues is larger than one, the flow in the radial direction cancels at a value $\rho_0$ given by Eq.~\eqref{scal_fp}, which yields
     \begin{equation}
         \sigma^{-1} = \left\langle \phi^\prime\left(0, \rho_0^2\right)\right\rangle.
     \end{equation}
     
     Based on Eq.~\eqref{rho}, we observe that any perturbation in the plane away from the limit cycle makes the radial component $\rho$ go back to $\rho_0$. The limit cycle is therefore stable, as shown in Fig.~\ref{fig:T2} G.
     
      Introducing this result into Eq.~\eqref{ang}, we obtain that the oscillations of the limit cycle are generated at a frequency 
     \begin{equation}\label{freq_LC}
         \omega_{LC} = \frac{\sigma_{\omega}}{\sigma}.
     \end{equation}
     
     In this analysis, Eq.~\eqref{freq_LC} is derived for the particular covariance matrix $\boldsymbol{\sigma_{mn}}$ in Eq.~\eqref{comp_corr}, which is the sum of an isotropic matrix (proportional to the identity) and an antisymmetric matrix. However, this equation is valid more generally for any connectivity matrix with a pair of complex eigenvalues (see Appendix~\ref{app:4}). When the covariance matrix is not antisymmetric but still has complex eigenvalues, the limit cycle is no longer a circle but resembles an ellipse, while the frequency of oscillation is still given by Eq.~\eqref{freq_LC}.
     
     Complex eigenvalues can be combined with real eigenvalues in networks with rank larger than two. The same stability properties are kept: only the fixed point or the limit cycle generated by the eigenvector with largest real part is stable. The unstable fixed points or limit cycles remain attractive within the dimensions spanned by the corresponding eigenvector (see Figure \ref{fig:SFig2} in Appendix~\ref{app:4}) for an example of a rank-three network combining an unstable limit cycle and two stable fixed points). 
     
\paragraph{Slow manifolds}
    When the covariance matrix $\boldsymbol{\sigma_{mn}}$ has degenerate eigenvalues, low-rank RNNs can lead to other phenomena than discrete fixed points or limit cycles. As an example of degenerate eigenvalues, we study the network dynamics when the covariance matrix $\boldsymbol{\sigma_{mn}}$ is diagonal:
    \begin{equation}\label{cov_ring}
        \boldsymbol{\sigma_{mn}} = \sigma_{mn}\boldsymbol{I}.
    \end{equation}
    \noindent This covariance matrix has one single real eigenvalue $\sigma_{mn}$, which is degenerate, since it has $R$ linearly independent eigenvectors. Introducing the covariance matrix in Eq.~\eqref{cov_ring} into the dynamics in Eq.~\eqref{dyn_vec} we obtain the fixed point equation
    \begin{equation}\label{fp_ring}
        \boldsymbol{\kappa}_0 = \left\langle \phi^\prime \left(0, \boldsymbol{\kappa}_0^T \boldsymbol{\kappa}_0\right) \right\rangle\sigma_{mn} \boldsymbol{\kappa}_0.
    \end{equation}
    \noindent To solve the fixed point equation, as in the previous section, we use the ansatz $\boldsymbol{\kappa}_0 = \rho_0 \boldsymbol{u_{\kappa_0}}$, where $\boldsymbol{u_{\kappa_0}}$ is an arbitrary unitary vector in collective space. Introducing the ansatz in the fixed point equation (Eq.~\ref{fp_ring}), we find that there is a non-trivial solution given implicitly by the scalar equation $\left\langle \phi^\prime \left(0, \rho_0^2\right) \right\rangle=\sigma_{mn}^{-1}$, which is independent of the particular direction $\boldsymbol{u_{\kappa_0}}$. Furthermore, we find that the fixed point is stable in the direction $\boldsymbol{u_{\kappa_0}}$. Therefore, in the mean-field limit given by Eq.~\eqref{dyn_vec}, this degenerate connectivity leads to a continuous manifold of attractive states that are at an equal distant $\rho_0$ away from the origin. In the case of rank-two connectivity, this degenerate covariance matrix leads to a stable ring attractor (Fig.~\ref{fig:T2} I-K), and in rank-$R$, to a stable $R$-spherical attractor. 
    
    In finite-size simulations, the  sampling of random loadings introduces spurious correlations in the matrix $\boldsymbol{\sigma_{mn}}$, breaking the degeneracy of the eigenvalues. As a consequence, only a small number of points on the  continuous attractor predicted by the mean-field theory give rise to actual fixed points. While the rest of the points on the predicted continuous attractor are not fixed points of the finite-size network, the dynamics around them are typically slow. More specifically, any trajectory of activity quickly converges towards the predicted continuous attractor, and then slowly evolves along it until it reaches a fixed point (Fig.~\ref{fig:T2} L) \citep{Mastrogiuseppe2018}. In finite-size networks, the continuous attractor predicted by the mean-field analysis therefore gives rise  to a low-dimensional manifold in state space, along which the dynamics are slow.

 When degenerate and non-degenerate real and complex eigenvalues are combined, the global stability appears to be given by the criterion in Eq.~\eqref{stab}: each eigenvalue generates its corresponding non-trivial dynamics (fixed points, continuous attractors or limit cycle) independently. The stability of these dynamical phenomena depends on the global eigenspectrum: the eigenvalues with the largest real part generate stable attractors, while the other eigenvalues lead to repellers. 
 
 \paragraph{Summary} In a low-rank network consisting of a single Gaussian population, the  possible non-trivial steady states are a pair of fixed points, a limit cycle, or a continuous attractor that gives rise to a small number of fixed points in finite networks. On top of these limited range of stable solutions, increasing the rank leads to additional unstable fixed points and limit cycles, that can potentially be used to control the dynamics, a point we do not further explore here. We instead proceed to show that increasing the number of Gaussian populations allows networks to implement a larger range of stable dynamics.

\FloatBarrier

\section{Dynamics in networks with multiple populations}

   As described in the previous section, a major limitation of rank-$R$ networks consisting of a single Gaussian population is that they cannot give rise to more than two stable fixed points, symmetrically arranged around the origin. We next show that networks consisting of several Gaussian populations can exhibit a larger number of stable fixed points.
    We specifically describe two different mechanisms by which multiple fixed points can be generated and controlled, and show that classical Hopfield networks \citep{Hopfield1982} correspond to a particular limit of Gaussian-mixture low-rank networks.
   
   \paragraph{Non-linear gain control}
    The first mechanism for generating multiple fixed points is based on having several populations that reach saturation in different regions of the collective space. The local dynamics in these different regions are then controlled solely by the statistics of the non-saturated populations. 
    
\begin{figure}[h]
     \centering
     \begin{subfigure}[b]{\textwidth}
        \includegraphics[width=\linewidth]{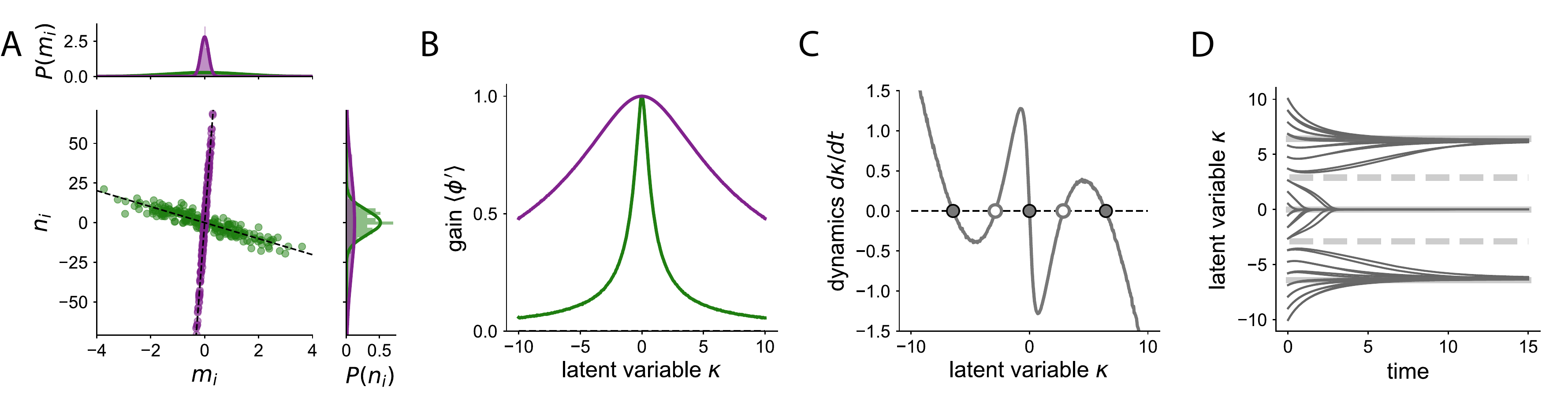}
     \end{subfigure}
	   	\caption{\textbf{Non-linear gain control in rank-one networks with multiple populations}. \textbf{A} Scatter plot between the components of the connectivity patterns $m_i$ and $n_i$ in a rank-one network with two Gaussian populations, shown in green (negatively correlated population) and purple (positively correlated population). \textbf{B} Average gain within the two populations (green and purple lines) at different states of the collective variable $\kappa$. The green population (large variance $\sigma_{m^2}^{\left(1\right)}$) saturates at values much closer to the origin than the purple population (low variance $\sigma_{m^2}^{\left(2\right)}$). Therefore, for large values of $\kappa$, the purple population has a stronger influence on the dynamics. \textbf{C}  Mean-field dynamics generated by the two-population statistics. Three stable fixed points (filled grey dots) emerge in the 1D recurrent dynamics. Close to the origin $\kappa_0$, the fixed point is stable because the green population dominates, and it has a negative correlation $\sigma_{mn}^{\left(1\right)}<0$. Therefore, the origin is stable. At values of $\kappa$ far from the origin, the purple population dominates, and it creates non-trivial stable fixed points. \textbf{D} Dynamics of the collective variable $\kappa$ in a network with $N=1000$ units, initiated at different initial values. The dynamics converge to one of the three stable fixed points. Parameters: $\sigma_{mn}^{\left(1\right)} = -10, \sigma_{mn}^{\left(2\right)} = 4.5, \sigma_{m^2}^{\left(1\right)}=1.98, \sigma_{m^2}^{\left(2\right)} = 0.02$, $\sigma_{n^2}^{\left(1\right)}=59.5, \sigma_{n^2}^{\left(2\right)} =1020$, and $\alpha_1 = \alpha_2 = 0.5$.} \label{fig:T3}
    \end{figure}

    For concreteness, we consider a rank-one network  consisting of two populations,   defined by different statistics of pattern loadings. Within population $p$, for $p=1,2$, the joint distribution of $n$ and $m$ values over neurons is specified by a $2 \times 2$ covariance matrix $\Sigma^{(p)}$. For simplicity we take the mean of the distribution to be zero. In the two-dimensional loading space defined by $m$ and $n$, the two populations correspond to different Gaussian clusters, both centered at zero but with different shape and orientations (green and purple dots in Fig.~\ref{fig:T3} A).

    The dynamics of the collective variable $\kappa$ in Eq.~\eqref{eq:circuit}  read:
   \begin{equation}\label{twopop}
       \tau \frac{d\kappa}{dt} = - \kappa + \tilde{\sigma}_{mn} \kappa,
   \end{equation}
   with the effective feedback $\tilde{\sigma}_{mn}$ defined as
   \begin{equation}\label{effective_feedback}
   \tilde{\sigma}_{mn}=\frac{1}{2}\sigma_{mn}^{\left(1\right)} \left\langle \phi^\prime\left(0, \kappa^2 \sigma_{m^2}^{\left(1\right)}\right)\right\rangle +\frac{1}{2}\sigma_{mn}^{\left(2\right)} \left\langle \phi^\prime\left(0, \kappa^2 \sigma_{m^2}^{\left(2\right)}\right)\right\rangle.
   \end{equation}
   This effective feedback $\tilde{\sigma}_{mn}$ is set by the average of covariances $\sigma_{mn}^{\left(p\right)}$ for each population $p$, weighted by the gain of the population. Low gain implies that the population is at a saturated state. The parameter $\sigma_{m^2}^{\left(p\right)}$ controls the range at which a population saturates as the collective variable $\kappa$ increases. If the two populations have different variances $\sigma_{m^2}^{\left(p\right)}$, their gains will vary differently with $\kappa$ (Fig.~\ref{fig:T3} B). If moreover the different populations have covariances $\sigma_{mn}^{\left(p\right)}$ of different signs, the total effective feedback will vary strongly at different ranges of $\kappa$, while this is not the case in networks with uniform populations or a single one. Therefore, by manipulating the variance of the $\mathbf{m}$ connectivity pattern within each population, and the overlap between the left and right connectivity patterns, it is possible to generate more flexible dynamics.
   
    In particular, the network can have three stable fixed points: one at the origin, and a pair of symmetrical non-trivial fixed points. First, the origin $\kappa=0$ is always a fixed point of dynamics in Eq.~\eqref{twopop}. The origin is moreover a stable fixed point if the effective feedback at zero, which is given by $\frac{1}{2}\left(\sigma_{mn}^{\left(1\right)}+\sigma_{mn}^{\left(2\right)}\right)$, is smaller than $1$. Second, at large values of $\kappa$ the effective feedback $\tilde{\sigma}_{mn}$ should be positive to cancel the contribution of the leaky term $-\kappa$ and generate a non-trivial fixed point.  Therefore, one of the populations, which we define to be the first one ($p=1$), must have a strong negative overlap, $\sigma_{mn}^{\left(1\right)}<2-\sigma_{mn}^{\left(2\right)}< 0$. Given Eq.~\ref{effective_feedback}, this implies that the gain of the positively correlated population two should be large, whereas the gain of the negatively correlated population one should be close to zero. A small gain is achieved in the first population by having a large value $\sigma_{m^2}^{\left(1\right)}$, so that the second condition reads $\sigma_{m^2}^{\left(1\right)}\gg \sigma_{m^2}^{\left(2\right)}$.  Fig.~\ref{fig:T3} C-D shows the dynamics of such a network given by the mean-field equation and in finite-size networks. 
   
   More generally, with more than two populations this mechanism  can be extended to produce a larger number of stable fixed points in rank-one networks. The two key components of this mechanism are: (i) an independent control of the gain of the different populations, so that the contribution of each population to the effective feedback takes place in different ranges of the collective variable $\kappa$; (ii) covariances $\sigma_{mn}^{\left(p\right)}$ of different signs, so that the effective feedback can flexibly take both positive and negative values in different ranges of $\kappa$. These mechanisms can also be applied to networks with rank higher than one. In that case, the overlap between loadings is given by a matrix $\boldsymbol{\sigma_{mn}^{\left(p\right)}}$ instead of a scalar, while the gain of each population is a scalar value. Populations with different covariance matrices and gains that vary at different ranges of the collective variables are able to generate multiple fixed points in different regions of the collective space, or combinations between stable limit cycles and stable fixed points \citep{Dubreuil2020}.
   
    \FloatBarrier
    \paragraph{Symmetries in loading space}
    A second mechanism for generating multiple fixed points is to exploit symmetries in the distribution of loadings $P\left(\underline{m}, \underline{n}\right)$. Such a symmetry in the connectivity induces  a symmetry in the dynamics of the collective variables. In consequence, if a network  generates one non-trivial stable fixed point, additional stable fixed points appear at  symmetric points in the collective space. 
    
    We focus here on networks where the overlap between the connectivity patterns is given by the non-zero means of the loadings, which is complementary to the previous section where the connectivity patterns had zero mean and the recurrent dynamics is determined by the covariances between the loadings. We introduce a symmetry in the distribution of loadings by arranging the means of the loadings accordingly. Note however that symmetrical distributions of loadings can also be generated in the zero-mean case.
    
    As an illustration, we consider first a rank-two network, with units evenly split into $P$ populations.  In each population, the loadings $m_1^{\left(p\right)}, m_2^{\left(p\right)}, n_1^{\left(p\right)}, n_2^{\left(p\right)}$ have a different set of means $a_{m_1}^{\left(p\right)}$, $a_{m_2}^{\left(p\right)}$, $a_{n_1}^{\left(p\right)}$, $a_{n_2}^{\left(p\right)}$ and the covariances $\sigma_{m_r n_s}^{\left(p\right)}$ are zero. The variance of the loadings, $\sigma_{m^2}$ and $\sigma_{n^2}$, are identical in all populations. As a consequence, different populations correspond to clusters of identical spherical shape, but centered at different points in the four-dimensional loading space. 
    
\begin{figure}[h]
     \centering
     \begin{subfigure}[b]{0.87\textwidth}
        \includegraphics[width=\linewidth]{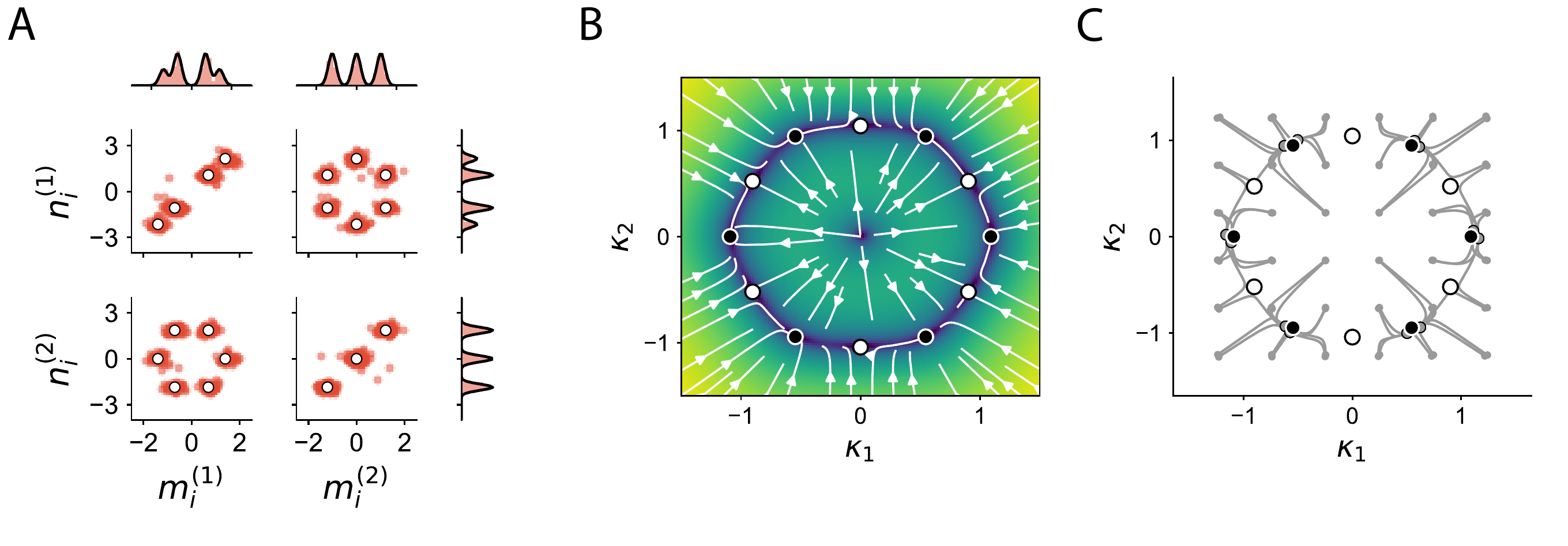}
     \end{subfigure}
	   	\caption{\textbf{Discrete rotational symmetry in rank-two networks with multiple populations}.  \textbf{A}  Scatter plot between the components of the connectivity patterns in a rank-two network. The network consists of six populations, with centers located on the vertices of a regular hexagon. The distribution is invariant to rotations of an angle $2\pi / 6$ in the $m_1$\textendash$n_2$ and $m_2$\textendash$n_1$ planes. \textbf{B} Mean-field dynamics of the network, the colormap represents the speed of the dynamics $Q$ (blue: slow dynamics, yellow: fast dynamics). The hexagonal symmetry in the loadings produces a solution with hexagonal symmetry, with six stable fixed points (black dots) symmetrically arranged along a ring. Saddle points (white dots) appear between the stable fixed points. \textbf{F} Trajectories of the collective variables in finite-size simulations, initiated at different initial conditions. All trajectories converge to one of the six stable fixed points. Two different network realizations are shown for each initial condition.  Parameters: centers arranged as in Eqs.~\eqref{centers1} and ~\eqref{centers2} where $p=6$ and $R_n=1.5$. Variance $\sigma_{n^2}=0.2$, equal in each population. Network size $N=1000$.} \label{fig:T4}
    \end{figure}

    We specifically arrange the means of the different populations (centers of the different clusters) symmetrically  at the vertices of a regular polygon in the planes of loadings $m_1-m_2$ and $n_1-n_2$:
    
    \begin{align}\label{centers1}
        a_{m_1}^{\left(p\right)} &= R_m \cos\left(\frac{2\pi p}{P}\right), \quad  a_{m_2}^{\left(p\right)} = R_m \sin\left(\frac{2\pi p}{P}\right);\\
        a_{n_1}^{\left(p\right)} &= R_n \cos\left(\frac{2\pi p}{P}\right), \quad  a_{n_2}^{\left(p\right)} = R_n \sin\left(\frac{2\pi p}{P}\right);\label{centers2}
    \end{align}
    \noindent where $p$ is the population index, $p=1\dots P$. The radial distance $R_m$ is fixed so that the patterns $\mathbf{m^{\left(1\right)}}$ and $\mathbf{m^{\left(2\right)}}$ have unit variance, while the free parameter $R_n$ controls the overlap between the connectivity patterns. Figure $\ref{fig:T4}$ A shows an example  with six populations, $P=6$. This distribution has a discrete rotational symmetry of order $P$, since rotations of angle $2\pi/P$ in the planes $m_1-n_2$ and $m_2-n_1$ leave the distribution unchanged. 
    
    Using the mean-field description in Eq.~\eqref{dyn_pop_general_cont}, the dynamics of the two collective variables now read
    \begin{align}\label{v_k1}
        \tau \frac{d\kappa_1}{dt}&=-\kappa_1 + \frac{1}{P}\sum_{p=1}^{P} a_{n_1}^{\left(p\right)} \left\langle \phi\left(a_{m_1}^{\left(p\right)} \kappa_1 + a_{m_2}^{\left(p\right)} \kappa_2, \sigma_m^2 \left(\kappa_1^2 + \kappa_2^2\right)  \right) \right\rangle\\
        \tau \frac{d\kappa_2}{dt}&=-\kappa_2 + \frac{1}{P}\sum_{p=1}^{P} a_{n_2}^{\left(p\right)}  \left\langle \phi\left(a_{m_1}^{\left(p\right)} \kappa_1 + a_{m_2}^{\left(p\right)} \kappa_2 ,\sigma_m^2 \left(\kappa_1^2 + \kappa_2^2\right) \right) \right\rangle\label{v_k2}.
    \end{align}
    
    Given the symmetry in the distribution, if we identify one non-trivial stable fixed point, there will be at least $P-1$ other fixed points with the same stability. Focusing on the direction given by $\kappa_2=0$, the velocity in the $\kappa_2$ direction, given by Eq.~\eqref{v_k2}, is always zero due to the symmetry in the distribution. Therefore, we  obtain a fixed point equation for $\kappa_1$ on the $\kappa_2=0$ direction using Eq.~\eqref{v_k1}:
    \begin{equation}
        \kappa_1 = \frac{1}{P} \sum_{p=1}^{P} R_n \cos\left(\frac{2\pi p}{P}\right) \left\langle \phi\left(R_m \cos\left(\frac{2\pi p}{P}\right) \kappa_1 , \sigma_m^2 \kappa_1^2  \right) \right\rangle.
    \end{equation}
    
    \noindent The r.h.s. is a sum of $P$ monotonically increasing bounded functions of $\kappa_1$. If the slope at the origin is larger than one, then, the r.h.s. will intersect with the function $\kappa_1$ at a non-trivial point. The slope of the r.h.s at the origin, obtained by differentiating the r.h.s. with respect to $\kappa_1$ and evaluating at  $\kappa_1=0$, is $\frac{1}{2}R_n R_m$, so that a condition for a non-trivial fixed point is 
    \begin{equation}
        R_n R_m >2.
    \end{equation}
    
\begin{figure}[h]
     \centering
     \begin{subfigure}[b]{0.9\textwidth}
        \includegraphics[width=\linewidth]{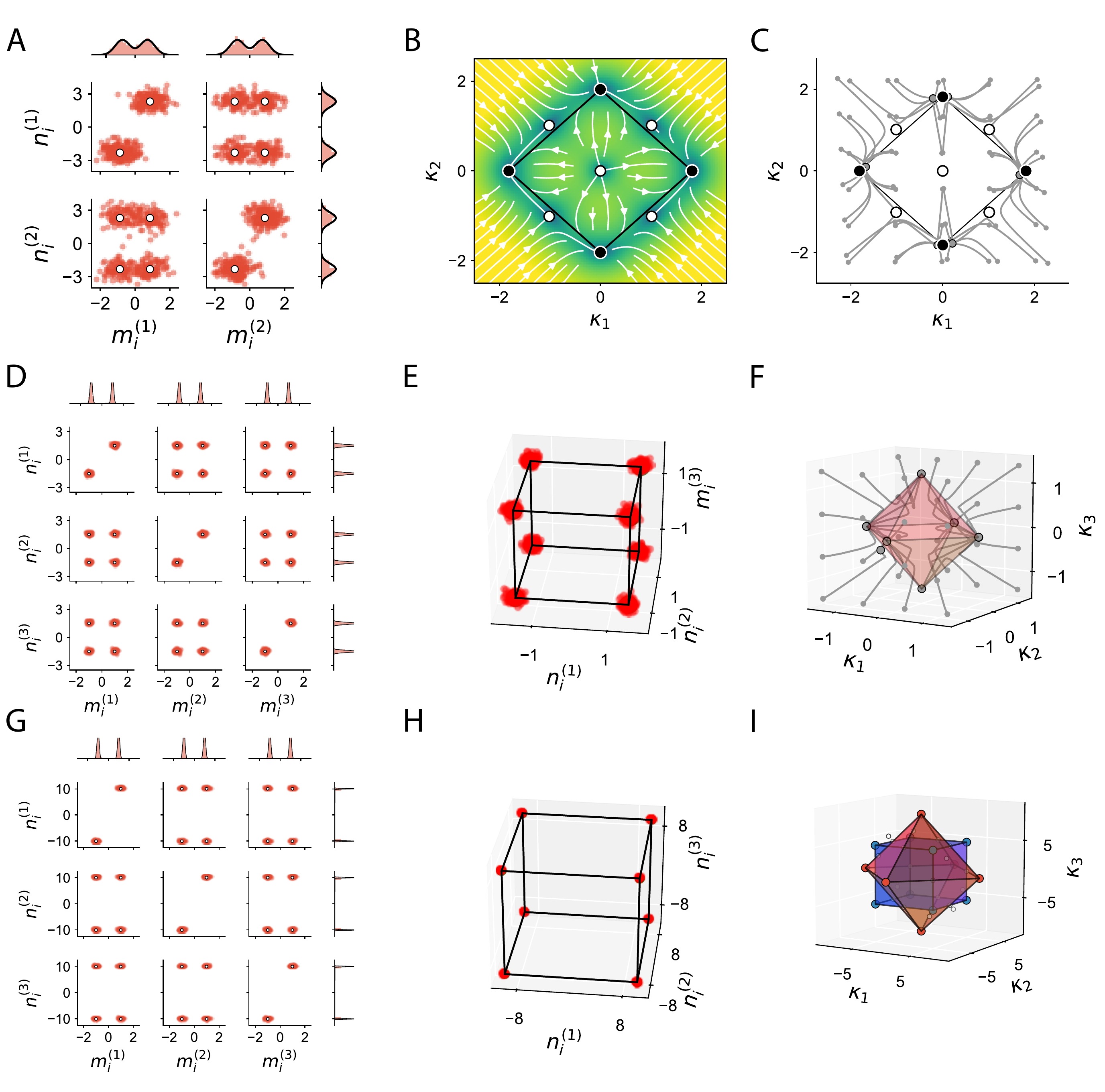}
     \end{subfigure}
	   	\caption{\textbf{Dynamics in rank-$R$ networks with discrete symmetry in multiple populations: Hopfield networks.} \textbf{A} Scatter plot between the entries of left singular vectors $m_i$ and right singular vectors $n_i$ in a rank-two network with four populations following Eqs.~\eqref{centers1} and \eqref{centers2}, with $P=2$, $R_n = 2.3$ and $\sigma_{n_r^2}=0.5$.  \textbf{B} Corresponding mean-field dynamics. The colormap represents the speed of the dynamics, defined as the norm of vector $\frac{d\boldsymbol{\kappa}}{dt}$ (blue: slow dynamics, yellow: fast dynamics). Four stable fixed points emerge, arranged in a square. \textbf{C} Trajectories starting at different initial conditions in a finite-size network. Each initial condition shows trajectories for two network realizations. \textbf{D} Analogous to  \textbf{A} in a rank-three network with loadings arranged as in Eqs.~\ref{cent_m} and \ref{cent_n}. \textbf{E} The populations are arranged at the vertices of a cube in loading space. $R_n = 2.1$. \textbf{F} Dynamics of the collective variables. Six stable fixed points (grey dots) emerge, arranged at the vertices of a dodecahedron (dual polygon of the cube, highlighted in red for visual purposes). Grey lines correspond to the trajectories of finite-size networks, initialized at different points in state-space.  \textbf{G-I} Same as in \textbf{D-F}, but for a network whose populations have larger mean values, $R_n=7$. For such large values, spurious fixed points that are proportional to the combinations of the three stored patterns $\left(\pm \boldsymbol{m}_1 \pm \boldsymbol{m}_2 \pm \boldsymbol{m}_3,\right)$ also become stable. Therefore, apart from the six fixed points in a octahedron (red polygon), eight other spurious fixed points appear arranged in a cube (blue polygon). Network size $N=1000$.} \label{fig:T5}
    \end{figure}

    Because of the symmetry, if $R_m R_n > 2$, there are at least $P$ stable fixed points arranged symmetrically on a circle (Fig.~\ref{fig:T4} A-C for a network with $P=6$ populations, and Fig.~\ref{fig:T5} A-C for a network with $P=4$ populations). If the number of population pairs is odd, there are $2P$ stable fixed points symmetrically arranged on a circle, because there is also a symmetry with respect to the origin, imposed by the symmetry in the transfer function. Otherwise, if $P$ is even, $P$ stable fixed points are generated by the network.
    
    Symmetrical arrangements of multiple populations can also be used  in higher $R$-rank networks to obtain multiple stable fixed points located on a $R$-dimensional sphere. For example, in rank-three networks, we consider eight populations whose centers are arranged at the vertices of a cube. The centers of the eight populations in the three-dimensional space of loadings $m^{\left(r\right)}$, for $r=1,2,3$, correspond to the vertices of a cube with side $2R_m$, so that
    \begin{align}\label{cent_m}
        \left(a_{m_1}^{\left(p\right)}, a_{m_2}^{\left(p\right)}, a_{m_3}^{\left(p\right)}\right) = \left(\pm R_m, \pm R_m, \pm R_m\right).
    \end{align}
    
    \noindent Populations $p = 1, \dots,8$ correspond to one of the eight different possible combinations of the sign. The variances of the loadings, $\sigma_{m^2}$ is identical in all populations. The value of $R_m$ is fixed so that the norm of each connectivity pattern $\mathbf{m^{\left(r\right)}}$ is $N$.
    
    The centers of the $n^{\left(r\right)}$ loadings follow the same configuration, at the vertices of a cube of side $2 R_n$:
    \begin{align}\label{cent_n}
        \left(a_{n_1}^{\left(p\right)}, a_{n_2}^{\left(p\right)}, a_{n_3}^{\left(p\right)}\right) = \left(\pm R_n, \pm R_n, \pm R_n\right),
    \end{align}
    \noindent where each population $p$ correspond to the same combination of signs as for the $m$ loadings, so that 
    \begin{equation}\label{prop}
        \text{sgn}\left(a_{m_r}^{\left(p\right)}\right) = \text{sgn}\left(a_{n_r}^{\left(p\right)}\right),
    \end{equation}
    \noindent with the collective index $r=1,2,3$  and the population index $p=1\dots 8$. The value $R_n$ is, as in the previous case, a free parameter that controls the overlap between connectivity patterns. This configuration is shown in Fig.~\ref{fig:T5}  D-E and G-H, for two different values of $R_n$. This distribution exhibits a cubic symmetry in the loading space $m_1-m_2-m_3$ and in space $n_1-n_2-n_3$. Thus, if we identify a non-trivial fixed point, these symmetries require the existence of symmetric solutions in the collective space. Inspecting the direction $\kappa_2=\kappa_3=0$ in the dynamics, we obtain a criterion for having a non-trivial stable fixed point:
    \begin{equation} \label{eq:rank3_sym}
        \kappa_1 = \frac{1}{8}\sum_{p=1}^8  a_{n_1}^{\left(p\right)}\left\langle\phi\left(a_{m_1}^{\left(p\right)}\kappa_1, \sigma_m^2 \kappa_1^2\right)\right\rangle
    \end{equation}
    \noindent Eq.~\ref{eq:rank3_sym} has a non-trivial solution, which is always stable, if $R_n R_m>1$.
    When this solution exists, applying a rotation of $\pi/2$ in the $m_1-m_2$ plane and in the $m_1-m_3$, it is possible to determine the other five stable fixed points that are generated by the symmetry (Fig.~\ref{fig:T5} F). These stable fixed points are arranged in the collective space at the vertices of an octahedron, the dual polyhedron of the cube (the dual of a polyhedron $A$ is the polyhedron $B$ where the vertices of $A$ correspond to the edges of $B$). Applying symmetry principles, the middle point of each triangular face of the octahedron is also a fixed point. However, the stability of this fixed point depends on the overlap $R_n R_m$. If $R_n R_m$ is larger than one but low, these fixed points are saddle points (Fig.~\ref{fig:T5} F). Beyond a critical value of $R_n R_m$, these fixed points become also stable. This second set of fixed points consists of eight points arranged on a cube (Fig.~\ref{fig:T5} I, blue dots). 
    
    In general, any $K$-dimensional discrete symmetry in the loadings will generate a dynamical system with stable fixed points on a $K$-dimensional sphere, arranged with the symmetry of the dual polytope. A regular polytope is defined as the generalization of a regular polyhedron generalized to more than three dimensions.
    
    \paragraph{Relation to Hopfield networks}
    
    Classical Hopfield networks \citep{Hopfield1982} storing $R\ll N$ patterns can be seen as a particular limit of  Gaussian-mixture low-rank networks, where  multiple stable fixed points are generated based on  symmetries in the connectivity. A Hopfield network is designed to store $R$ binary patterns $m_i^{\left(r\right)} = \pm m$, where for every neuron the sign of the entry in each pattern generated randomly, and $m$ is a scalar parameter.
    A Hopfield network storing these $R$ patterns is defined as a recurrent network with connectivity matrix
    
    \begin{equation}\label{eq:hopfield}
        J_{ij}^{Hopfield} = \sum_{r=1}^R m_i^{\left(r\right)} m_j^{\left(r\right)} 
    \end{equation}
    
    \noindent  Such a configuration  corresponds to a specific type of low-rank matrix, and can be mapped onto Gaussian-mixture low-rank networks. A first specific property of Hopfield networks (Eq.~\ref{eq:hopfield}) is that the connectivity is symmetric, so that the left and right connectivity patterns are proportional to each other
    \begin{equation}
    \mathbf{m^{\left(r\right)}} = c \mathbf{n^{\left(r\right)}}    
    \end{equation}
    \noindent where $c$ is a positive constant. A second specific property is that the loadings of the patterns $\mathbf{m^{\left(r\right)}}$ and $\mathbf{n^{\left(r\right)}}$, for $r=1,\dots,R$, are binary and of equal sign, so that each neuron is characterized by $2R$ loadings that can only differ from each other in their signs. Therefore, each neuron in a Hopfield network belongs to one of the $2^R$ sign combinations allowed. In terms of the low-rank framework, Hopfield networks can therefore be described as low-rank networks with $2^R$ deterministic populations, which have means 

    \begin{align}\label{cent_mR}
        \left(a_{m_1}^{\left(p\right)}, \cdots, a_{m_R}^{\left(p\right)}\right) &= R_m \left(\pm 1, \dots, \pm 1\right),\\
        \left(a_{n_1}^{\left(p\right)}, \cdots, a_{n_R}^{\left(p\right)}\right) &= R_n \left(\pm 1, \dots, \pm 1\right),\\
        \text{sgn}\left(a_{m_r}^{\left(p\right)}\right)&=\text{sgn}\left(a_{n_r}^{\left(p\right)}\right),
    \end{align}
    
    \noindent and where there is no dispersion around the mean of each population, so that $\sigma_{m}^{\left(p\right)}=\sigma_{n}^{\left(p\right)}=0$.

A rank-two network with four populations $P=4$ \textendash\,characterized by Eq.~\eqref{centers1}, see Fig.~\ref{fig:T5} A-C,\textendash\, is therefore equivalent to a two-pattern Hopfield network in the limit of no dispersion around the mean of each cluster, $\sigma_{m^{2}}^{\left(p\right)} = 0$. In this limit, saddle points are located at the midpoints between  neighbouring stable fixed points. In the more general rank-two networks in Eq.~\eqref{centers1} where $\sigma_{m^{2}}^{\left(p\right)} > 0$, the saddle points between stable fixed points move further away from the origin (such as in Fig.~\ref{fig:T5} B, where $\sigma_{m^{2}}^{\left(p\right)}=0.3$), but the four stable fixed points remain on the vertices of a square along the axes $\kappa_1 = 0$ and $\kappa_2 = 0$. In the limit of very large $\sigma_{m^{2}}^{\left(p\right)}$ the saddle points between stable fixed points approach the circle that circumscribes the stable fixed points. 

The rank-three network presented in Eqs.~\eqref{cent_n} and \eqref{prop} also becomes a classical Hopfield network in the limit of $\sigma_{m^{2}}^{\left(p\right)} \to 0$. Allowing for values $\sigma_{m^{2}}^{\left(p\right)} > 0$, as illustrated in Fig.~\ref{fig:T5} D and G, does not change the number of fixed points generated by the Hopfield network nor their direction in collective space. These networks generate pairs of stable fixed points along the directions $\mathbf{m^{\left(1\right)}}$, $\mathbf{m^{\left(2\right)}}$, and $\mathbf{m^{\left(3\right)}}$. When $R_m R_n$ is large, 
 additional  fixed points become stable along directions $\pm \mathbf{m^{\left(1\right)}} \pm \mathbf{m^{\left(2\right)}} \pm \mathbf{m^{\left(3\right)}}$. These additional fixed points correspond to well known spurious mixture states in Hopfield networks \citep{Amit1987}.

\FloatBarrier

\section{Approximating dynamical systems with Gaussian-mixture low-rank networks}
    In the previous section, we focused on generating multiple fixed points in an autonomous network by means of a few Gaussian populations in the connectivity. More generally, as shown in Section 2.3, multi-population rank-$R$ networks can approximate any $R$-dimensional dynamical system. In this section, we propose an algorithm for that purpose.

     Previous works have developed algorithms for training recurrent networks to implement given dynamics that effectively used low-rank connectivity \citep{Eliasmith2003, Rivkind2017,  Pollock2019}. These methods rely on  tuning the loadings $n_i^{\left(r\right)}$ of individual neurons,  given fixed external inputs  $I_i^{\left(s\right)}$ and connectivity loadings $m_i^{\left(r\right)}$. Here we focus instead on networks based on mixtures of Gaussian populations, in which the couplings between individual neurons are not precisely set, but instead sampled from a distribution.  We extend previous methods to find the first and second order moments of multiple Gaussian populations that approximate a given dynamical system.
    
      Our goal is to approximate the $R$-dimensional dynamics specified by a vector field $G\left(\boldsymbol{\kappa}\right)$:
    \begin{equation}\label{target}
        \frac{d\boldsymbol{\kappa}}{dt} =  G\left(\boldsymbol{\kappa}\right).
    \end{equation}
    
    Our algorithm proceeds as follows.  We first fix the number of Gaussian populations in the network and the fraction of neurons included in each population, $\alpha_p$. Depending on the complexity of the target dynamics and the required accuracy, a smaller or larger number of populations is required. Second, we set the mean and variance of the $\bf{m}^{\left(r\right)}$ vectors in each population, $a_{m_r}^{\left(p\right)}$ and $\sigma_{m_r^2}^{\left(p\right)}$, together with the mean and variance of the external input, $a_I^{\left(p\right)}$ and $\sigma_{I^2}^{\left(p\right)}$. We randomly assign these parameters randomly according to a certain probability distribution. Finally, we determine the statistics of the $\bf{n}^{\left(r\right)}$  vectors, the only unknown in the network, using linear regression.
    
    \noindent  We define a number of set points $\left\{\boldsymbol{\kappa}_k\right\}_{k=1\dots K}$ on which we impose that the effective flow in the low-rank network given by Eq.~\eqref{dyn_pop_general_cont} be equal to the target vector field
    \begin{equation}\label{EF_1}
         G\left(\boldsymbol{\kappa}_k\right)= -\boldsymbol{\kappa}_k + \sum_{p=1}^{P} \alpha_p \left(\boldsymbol{a_n^{(p)}} \left\langle \phi\left(\mu^{\left(p\right)}\left(\boldsymbol{\kappa}_k\right), \Delta^{\left(p\right)}\left(\boldsymbol{\kappa}_k\right)\right)\right\rangle + \boldsymbol{\sigma_{nm}^{\left(p\right)}} \boldsymbol{\kappa_k} \left\langle \phi^\prime\left(\mu^{\left(p\right)}\left(\boldsymbol{\kappa}_k\right), \Delta^{\left(p\right)}\left(\boldsymbol{\kappa}_k\right)\right)\right\rangle\right) .
    \end{equation}
    \noindent These $k=1\dots K$ set points should be relevant points of the vector field $G\left(\boldsymbol{\kappa}\right)$; they can be fixed points, but can also be chosen within a grid in collective space or based on sampled trajectories of the target system (Eq.~\ref{target}). For simplicity, in Eq.~\eqref{EF_1} we are considering that the input pattern $\bf{I}$ is orthogonal to the connectivity patterns $\bf{n}^{\left(r\right)}$. It is straight-forward to extend the algorithm to account for non-zero values of the parameters  $\sigma_{n_r I}$.
    
    Note that $\mu^{\left(p\right)}$ and $\Delta^{\left(p\right)}$ depend on the statistics of patterns $\bf{I}$ and $\bf{m}^{\left(r\right)}$ that are fixed (see Eq.~\eqref{eq:mu-delta}), but not on $a_{n_r}^{\left(p\right)}$ and $\sigma_{m_r n_r}^{\left(p\right)}$ which we aim to determine. Eq.~\eqref{EF_1} can therefore be written as a linear system of the form
    \begin{equation}
        \boldsymbol{G} = \boldsymbol{W}^T\mathbf{X}
    \end{equation}
    \noindent where, for each individual set point, $\boldsymbol{G}$ is a vector of length $R$, $\boldsymbol{G} = G\left(\boldsymbol{\kappa_{k}} \right) +\boldsymbol{\kappa_{k}}$,  the vector 
    \begin{equation}
    \boldsymbol{X} := \left[a_{n_1}^{\left(1\right)},\ldots a_{n_R}^{\left(1\right)}, \sigma_{m_1 n_1}^{\left(1\right)}, \dots \sigma_{m_1 n_R}^{\left(1\right)}, \dots \sigma_{m_R n_1}^{\left(1\right)}, \dots \sigma_{m_R n_R}^{\left(1\right)}, \dots a_{n_1}^{\left(P\right)} \dots \sigma_{m_R n_R}^{\left(P\right)} \right]
    \end{equation} 
    \noindent has length $R\left(R+1\right)P$ and the corresponding matrix $\boldsymbol{W}$ of size $ R \left(R+1\right) P \times R$. For the $K$ set points $\boldsymbol{\kappa}_k$ on which we want to approximate the dynamics, we concatenate the vector $\boldsymbol{G}$ and matrix $W$ of each point, so that they will be of size $R\cdot K$ and $R\cdot K \times \left(R\left(R+1\right) P \right)$ respectively. 
    
    The unknown values of vector $\mathbf{X}$ can now be obtained by standard linear regression as 
    \begin{equation}
        \boldsymbol{X} = \left(\boldsymbol{W}\boldsymbol{W}^T\right)^{-1}\boldsymbol{W} \boldsymbol{G}.
    \end{equation}
    
    \noindent Often, it is convenient to regularize the regression algorithm to avoid the entries of $\boldsymbol{X}$ being exceedingly large, at the cost of increasing the error in the approximation of the dynamics. Solutions with very large values of $\boldsymbol{X}$ are less robust, because they produce stronger finite-size effects when sampling from the found mixture of Gaussians, potentially affecting the stability of the solution. One standard possibility amongst many is to use ridge regression to find the unknown values
    \begin{equation}\label{regress}
        \boldsymbol{X} = \left(\boldsymbol{W}\boldsymbol{W}^T + \beta^2 \boldsymbol{I}\right)^{-1}\boldsymbol{W} \boldsymbol{G}
    \end{equation}
    \noindent where $\beta$ is the ridge parameter that controls the amount of regularization.
    
    The number of populations, the level of regularization, together with the distributions chosen to fix the mean and covariance values  $a_{m_r}^{\left(p\right)}$, $\sigma_{m_r^2}^{\left(p\right)}$, $a_I^{\left(p\right)}$ and $\sigma_{I^2}^{\left(p\right)}$ are hyperparameters of the algorithm. These hyperparameters can be tuned progressively by running several iterations of the algorithm. For example, a possible goal is to search for the the minimal number of populations required for approximating a given dynamical system within some accuracy limits. In general, we observe empirically that the distribution underlying the fixed parameters does not play a crucial role in the accuracy of the algorithm, as long as they span a wide range of values.
    
\begin{figure}[h]
     \centering
     \begin{subfigure}[b]{\textwidth}
        \includegraphics[width=\linewidth]{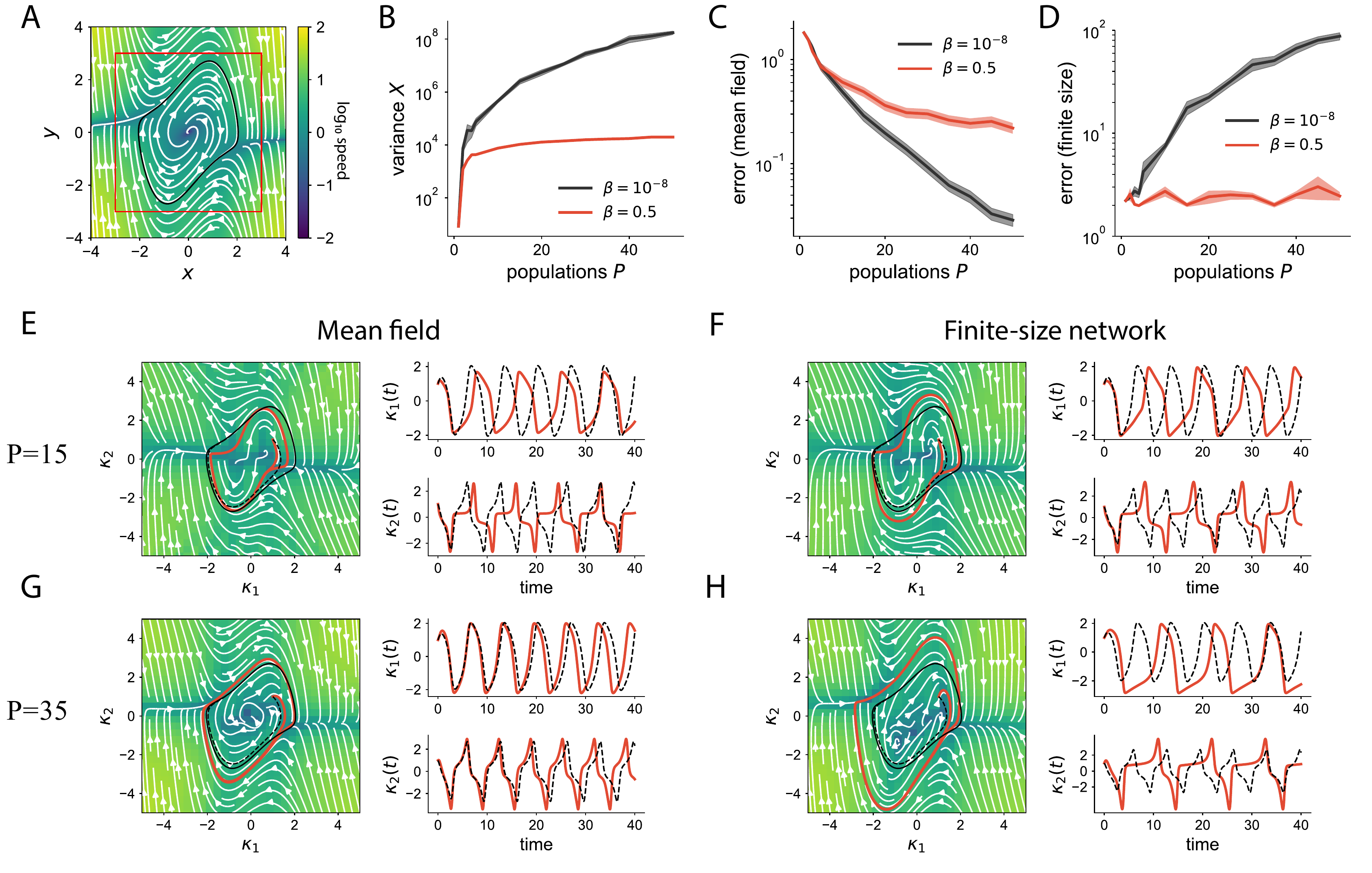}
     \end{subfigure}
	   	\caption{\textbf{Approximation of a Van der Pol oscillation with multiple-population low-rank networks}. \textbf{A} Dynamics of a Van der Pol oscillator ($\mu=1$). The red square indicates the boundaries of the grid used to approximate the dynamical system. \textbf{B} Variance of the obtained parameters $X$ for two different levels of regularization $\beta$ (Eq.~\ref{regress}. The variance grows extremely fast as the number of populations $P$ is increased in non-regularized networks. Error bars represent the standard error of the mean (SEM) with 50 realizations. \textbf{C} Average approximation error by the mean-field description as a function of the number of populations. The error is measured as the average deviation from the target vector field, assessed at the grid points used for training the algorithm. \textbf{D} Average approximation error by a finite-size network, with $N=2000$ units per population. The finite-size error quickly grows with no regularization. \textbf{E} Mean-field approximation with $P=15$ populations. Left: dynamical landscape. The red curve corresponds to a trajectory initiated at state $\left(1,1\right)$. In black, the corresponding trajectory of the Van der Pol oscillator. Right: trajectories as a function of time. \textbf{F} Finite-size network corresponding to the mean-field solution found in \textbf{E}. \textbf{G - H} Similar to \textbf{G-H}, with a larger number of populations, $P=35$. The mean-field solution approximates better the Van der Pol oscillators. However, the finite-size approximation does not improve, given that the number of units per population was kept constant ($N=2000$). Parameters: Solutions in $\textbf{E-H}$ are regularized, $\beta=0.5$. Values $\sigma_{m_r^2}^{\left(p\right)}$ and $\sigma_{I^2}^{\left(p\right)}$ are initially drawn from an exponential distribution with unit variance, $a_I^{\left(p\right)}$ and $a_{m_1}^{\left(p\right)}$ are drawn from a uniform distribution with bounds -2 and 2, and $a_{m_2}^{\left(p\right)}=0$.} \label{fig:T7}
    \end{figure}

    To illustrate the algorithm, we use a rank-two network to approximate a Van der Pol oscillator. The Van der Pol oscillator is a two-dimensional non-linear dynamical system that generates non-harmonic oscillations. It is defined as
    \begin{align}
        \frac{dx}{dt} &= y\\
        \frac{dy}{dt} &=  \mu\left(1-x^2\right)y-x
    \end{align}
    \noindent where $\mu$ is a scalar parameter that controls the strength of the non-linearity. For this example, we set $\mu=1$ (Fig.~\ref{fig:T7} A). Secondly, we determine the statistics of the left connectivity patterns and the external input, by drawing random values for the mean values in each population $a_I^{\left(p\right)}$ and $a_{m_r}^{\left(p\right)}$ from a zero-mean uniform distribution, and the variances $\sigma_{m_r^2}^{\left(p\right)}$ and $\sigma_{I^2}^{\left(p\right)}$ from an exponential distribution, all values of order one. As set points, we use a $K=30\times30$ grid for values $x$ and $y$ ranging between -3 and 3 (red square, Fig.~\ref{fig:T7} A). 
    
    We first analyze the performance of the algorithm as a function of the number of populations, for two different levels of regularization. We find that regularization is necessary to avoid diverging values of the parameters in $X$ (Fig.~\ref{fig:T7} B, black vs red curve). This is due to the fact that the matrix $\mathbf{\boldsymbol{W}\boldsymbol{W}^T}$ is close to singular, so that its inverse reaches very high values. Interestingly, the error in the approximation increases only slightly with a strong level of regularization (Fig.~\ref{fig:T7} C). As the number of populations is increased, the error in the approximation for all levels of regularization monotonically decreases. 
    
    The algorithm is based on the mean-field description in Eq.\eqref{EF_1}, which holds in the limit of very large number of neurons per population. We next study whether the obtained mean-field networks describe well the dynamics in networks with a finite number of neurons per population. We therefore sampled the connectivity and input loadings from the multivariate Gaussian distribution that characterizes each population. For that purpose, it is necessary to set the variances of the right connectivity patterns $\sigma_{n_r^2}^{\left(p\right)}$, which do not influence the mean field dynamics. As a general approach, we set those variances parameters to be as low as possible, but high enough so that the correlation matrix of the multivariate Gaussian distribution is positive-definite. Sampling a finite number of values from the distribution introduces deviations in the sampled mean and sampled covariance matrix that introduce additional finite-size errors in the approximated dynamics. We find that when the parameters obtained in $X$ reach very large values, the approximation error in networks with $N=2000$ per population is very large (Fig.~\ref{fig:T7} D, black curve). However, when the algorithm is constrained by a strong level of regularization, the finite-size error remain approximately constant as the population size increases (Fig.~\ref{fig:T7} D, red curve).
    
    We then show the approximated dynamics for two different number of populations, $P=15$ (Fig.~\ref{fig:T7} E-F) and $P=35$ (Fig.~\ref{fig:T7} G-H), in the regularized case. Fifteen populations are enough to obtain a limit cycle with similar features to the Van der Pol oscillator, although there are deviations in the shape and frequency of oscillation (Fig.~\ref{fig:T7} E). The finite-size effects remain small, so that the dynamics of a simulated network resemble the mean-field approximation (Fig.~\ref{fig:T7} F). For a larger number of populations, the mean-field approximation increases considerably the accuracy (Fig.~\ref{fig:T7} G). However, finite-size effects increase with the number of populations, when the number of units per population is kept constant, so that the approximation error in finite-size simulations is not necessarily reduced (Fig.~\ref{fig:T7} H).

    This algorithm can be applied to generate any given dynamics in collective space within a finite domain. Beyond this finite domain sampled through the chosen set points, if the target vector field does not follow the required asymptotic behavior (Eq.~\ref{leaky}), as it is the case for the Van der Pol oscillator, the network will not extrapolate to the target dynamics (region outside square $\left\lvert \kappa_1, \kappa_2\right\rvert>3$ in Fig.~\ref{fig:T7} E-H, left). However, in practice, it may produce qualitatively similar dynamics: in the example of the Van der Pol oscillator, if the network is initialized at a point outside the limit cycle, the resulting trajectories still converge to the limit cycle.

    \FloatBarrier
    \newpage
    
\section{Discussion}
In this manuscript, we have examined the dynamics in Gaussian-mixture low-rank recurrent neural networks, a class of models in which the connectivity is defined by a low-rank matrix, with connectivity patterns consisting of several populations with distinct Gaussian statistics. In these networks, the collective dynamics can be described by $R+N_{in}$ collective variables, where $R$ is the rank of the connectivity matrix and $N_{in}$ the dimensionality of the input patterns. These collective variables form a dynamical system, the evolution of which is determined by the connectivity statistics of the populations forming the network. The rank of the network, and the population structure therefore play complementary roles: the rank of the network sets the internal dimensionality of the dynamics and defines the corresponding collective variables, while individual populations  shape the dynamics of these collective variables, but do not contribute new ones. We specifically showed that, in the limit of a large number of populations, this class of network displays a universal approximation property, and can therefore implement a large range of dynamical systems. Having a small number of populations instead imposes constraints and limits the achievable range of dynamics.

We focused here on a specific family of distributions for the connectivity patterns, mixtures of multi-variate Gaussians. This choice was motivated by several considerations. First, this family of distributions can be used to approximate any multi-variate distribution for the pattern loadings. Second, this family of distributions leads to a particularly simple form of dynamics for the collective variables, where the time-evolution is formulated in terms of a simple effective circuit (Eq.~\ref{eq:circuit}). Remarkably, in this description of the dynamics, which is exact and  non-linear, the collective variables appear to interact linearly through effective couplings and effective inputs. This  allows for a particularly transparent interpretation of dynamics in terms of gain modulation. Several of our results are however independent of the specific assumption for the type of distribution; this is in particular the case for the influence of symmetry in the connectivity on the dynamics. When a large number of populations is needed to approximate the connectivity structure, other parametric distributions may be more suitable, and the interpretation in terms of discrete populations may not be appropriate.

Previously published studies \citep{Mastrogiuseppe2018, Mastrogiuseppe2019, Schuessler2020} have established the link between low-dimensional dynamics and low-rank networks, by focusing on fixed points and limit cycles in single population networks. In these studies, the connectivity matrix contained a random full-rank term and a low-rank random term. The full-rank term in the connectivity does not allow to express the emerging dynamics directly  as a low-dimensional dynamical system, as presented in this work (Section~\ref{section:model}), and to study the potential constraints on the low-dimensional dynamics. 

Low-rank networks with arbitrary pattern distributions form a rich and versatile framework that encompasses a number of previously studied types of recurrent neural networks. As shown in the last part of the results, Hopfield networks storing $R\ll N$ patterns can be seen as a particular limit of Gaussian-mixture low-rank networks, in which pattern loadings are binary and exhibit a specific type of symmetry. The Neural Engineering Framework \citep{Eliasmith2003} and the Manifold Embedding approach \citep{Pollock2019} provide algorithms that implement specific low-dimensional dynamics by controlling the structure of fixed points and Jacobians using linear-regression methods. These algorithms generate recurrent networks with low-rank connectivity, in which the individual couplings between individual neurons are set to specific values. In contrast, we focus on low-rank networks in which individual couplings are sampled from an underlying distribution, and our algorithm determines the statistics of this distribution rather than individual couplings. This approach automatically endows our networks with strong robustness with respect to deletion of individual neurons.


The dynamics of recurrent networks of rate units that include several populations have been studied in recent years \citep{Aljadeff2015a, Aljadeff2016, Kadmon2015}. In these works, the connectivity consisted of full-rank, random matrices in which populations define the pair-wise connectivity statistics.
 For example, in a two-population model of excitatory and inhibitory neurons, the inhibitory population is defined by the connectivity statistics (mean and variance of synaptic strengths) between inhibitory neurons and the connectivity statistics towards excitatory neurons. In contrast, here we focus on low-rank structure in the connectivity matrices, and specify populations in terms of the statistics for the low-rank structure vectors within each population. This allows us to go beyond random connectivity, and study the effects of the rank and the number of populations independently. Directly relating the two descriptions of connectivity is a topic of ongoing work.

Our framework is also closely related to Echo-state \citep{Jaeger} and FORCE networks \citep{Sussillo2009}, which rely on randomly connected recurrent networks controlled by feedback loops. Each feedback loop is mathematically equivalent to adding a unit-rank component to the connectivity matrix. Echo-state and FORCE networks  therefore correspond to low-rank networks with an additionnal full-rank, random term in the connectivity \citep{Mastrogiuseppe2018,Mastrogiuseppe2019}. Because the feedback loops are trained to produce specific outputs, the low-rank part of the connectivity is typically correlated to the random connectivity term (but see \cite{Mastrogiuseppe2019}). Such correlations increase the dimensionality and the  range of the dynamics \citep{Schuessler2020, Logiaco2019}, although the low-rank connectivity structure and the number of populations still generate strong constraints. For instance, for rank-one networks with a random term in the connectivity, but consisting of a single population, the fixed points are restricted to lie on a one-dimensional, but non-linear manifold, and typically at most two non-trivial stable fixed points can be generated \citep{Schuessler2020}. More generally, random components in the connectivity can strongly influence  learning dynamics during training \citep{Schuessler2020a}. 

Gaussian-mixture low-rank networks, the Neural Engineering Framework, and Echo-state networks all exhibit universal approximation properties \citep{Eliasmith2005, Maass2002}. It is however important to distinguish between several variants of this property. In our case, in analogy with the NEF, we  started from an $R$-dimensional dynamical system fully specified by its flow function, and showed that Gaussian-mixture low-rank networks can approximate this flow function, provided a large number of populations is available and the flow function satisfied specific constraints. Echo-state and FORCE networks instead start by specifying a target readout, and universal approximation means that any such readout can be generated by training the feedback \citep{Maass2007}. This readout corresponds to a low-dimensional projection of a large dynamical system, and Echo-state networks are free to implement any dynamical system consistent with the specified output projection. This is a major distinction with our, and the NEF approach, where the overall dynamical system is  more tightly constrained.

In this work, we have examined only networks with fixed inputs. Varying the inputs instead modifies the low-dimensional dynamics, an effect that can be understood through modulations of effective couplings that govern the interactions between collective variables. In a companion paper \citep{Dubreuil2020}, we have used Gaussian-mixture low-rank RNNs to reverse-engineer networks trained on a range of neuroscience tasks, and found that gain modulation through input control underlies complex computations, such as flexible input-output mappings \citep{Fusi2016}. Varying inputs while keeping connectivity fixed therefore has the potential of implementing a large range of dynamical systems and computations \citep{Pollock2019}, but the full capacity of this mechanism still remains to be fully elucidated.

\section*{Acknowledgements}
The project was supported by the Ecole de Neurosciences de Paris, the ANR project MORSE (ANR-16-CE37-0016), the CRCNS project PIND, the program “Ecoles Universitaires de Recherche” launched by the French Government and implemented by the ANR, with the reference ANR-17-EURE-0017.
There are no competing interests. We thank Mehrdad Jazayeri and Eli Pollock for discussions.

\section*{Code availability}
Code and trained models will be made available upon publication.

\newpage
\bibliography{universalapprox}

\begin{thebibliography}{}

\bibitem[Aljadeff et~al., 2016]{Aljadeff2016}
Aljadeff, J., Renfrew, D., Vegu{\'e}, M., and Sharpee, T.~O. (2016).
\newblock Low-dimensional dynamics of structured random networks.
\newblock {\em Physical Review E}, 93(2):022302.

\bibitem[Aljadeff et~al., 2015]{Aljadeff2015a}
Aljadeff, J., Stern, M., and Sharpee, T. (2015).
\newblock {Transition to chaos in random networks with cell-type-specific
  connectivity}.
\newblock {\em Physical Review Letters}, 114(8).

\bibitem[Amit et~al., 1987]{Amit1987}
Amit, D.~J., Gutfreund, H., and Sompolinsky, H. (1987).
\newblock {Statistical mechanics of neural networks near saturation}.
\newblock {\em Annals of Physics}, 173(1):30--67.

\bibitem[Barak, 2017]{Barak2017}
Barak, O. (2017).
\newblock {Recurrent neural networks as versatile tools of neuroscience
  research}.
\newblock {\em Current Opinion in Neurobiology}, 46:1--6.

\bibitem[Buonomano and Maass, 2009]{Buonomano2009}
Buonomano, D.~V. and Maass, W. (2009).
\newblock {State-dependent computations: Spatiotemporal processing in cortical
  networks}.
\newblock {\em Nature Reviews Neuroscience}, 10(2):113--125.

\bibitem[Chaisangmongkon et~al., 2017]{Chaisangmongkon2017}
Chaisangmongkon, W., Swaminathan, S.~K., Freedman, D.~J., and Wang, X.~J.
  (2017).
\newblock {Computing by Robust Transience: How the Fronto-Parietal Network
  Performs Sequential, Category-Based Decisions}.
\newblock {\em Neuron}, 93(6):1504--1517.e4.

\bibitem[Churchland and Shenoy, 2007]{Churchland2007}
Churchland, M.~M. and Shenoy, K.~V. (2007).
\newblock {Temporal complexity and heterogeneity of single-neuron activity in
  premotor and motor cortex}.
\newblock {\em Journal of Neurophysiology}, 97(6):4235--4257.

\bibitem[Cybenko, 1989]{Cybenko1989}
Cybenko, G. (1989).
\newblock {Approximation by superpositions of a sigmoidal function}.
\newblock {\em Mathematics of Control, Signals, and Systems}, 2(4):303--314.

\bibitem[Doya, 1993]{Doya1993}
Doya, K. (1993).
\newblock {Universality of Fully-Connected Recurrent Neural Networks}.
\newblock {\em Dept. of Biology, UCSD, Tech. Rep}, 1:1--6.

\bibitem[Dubreuil et~al., 2020]{Dubreuil2020}
Dubreuil, A., Valente, A., Beiran, M., Mastrogiuseppe, F., and Ostojic, S.
  (2020).
\newblock {Complementary roles of dimensionality and population structure in
  neural computations}.
\newblock {\em bioRxiv}, page 2020.07.03.185942.

\bibitem[Eliasmith, 2005]{Eliasmith2005}
Eliasmith, C. (2005).
\newblock {A unified approach to building and controlling spiking attractor
  networks}.
\newblock {\em Neural Computation}, 17(6):1276--1314.

\bibitem[Eliasmith and Anderson, 2003]{Eliasmith2003}
Eliasmith, C. and Anderson, C. H. C.~H. (2003).
\newblock {\em {Neural engineering : computation, representation, and dynamics
  in neurobiological systems}}.
\newblock MIT Press.

\bibitem[Fan et~al., 2020]{Fan2020}
Fan, F., Xiong, J., and Wang, G. (2020).
\newblock {Universal approximation with quadratic deep networks}.
\newblock {\em Neural Networks}, 124:383--392.

\bibitem[Funahashi, 1989]{Funahashi1989}
Funahashi, K.-I. (1989).
\newblock {On the approximate realization of continuous mappings by neural
  networks}.
\newblock {\em Neural Networks}, 2(3):183--192.

\bibitem[Fusi et~al., 2016]{Fusi2016}
Fusi, S., Miller, E.~K., and Rigotti, M. (2016).
\newblock {Why neurons mix: High dimensionality for higher cognition}.
\newblock {\em Current Opinion in Neurobiology}, 37:66--74.

\bibitem[Gallant and White, 1988]{Gallant1988}
Gallant, A.~R. and White, H. (1988).
\newblock There exists a neural network that does not make avoidable mistakes.
\newblock In {\em ICNN}, pages 657--664.

\bibitem[Gallego et~al., 2018]{Gallego2018}
Gallego, J.~A., Perich, M.~G., Naufel, S.~N., Ethier, C., Solla, S.~A., and
  Miller, L.~E. (2018).
\newblock {Cortical population activity within a preserved neural manifold
  underlies multiple motor behaviors}.
\newblock {\em Nature Communications}, 9(1):1--13.

\bibitem[Gao et~al., 2015]{Gao2015}
Gao, P., Ganguli, S., Battaglia, F.~P., and Schnitzer, M.~J. (2015).
\newblock {On simplicity and complexity in the brave new world of large-scale
  neuroscience This review comes from a themed issue on Large-scale recording
  technology}.
\newblock {\em Current Opinion in Neurobiology}, 32:148--155.

\bibitem[Hennequin et~al., 2014]{Hennequin2014}
Hennequin, G., Vogels, T.~P., and Gerstner, W. (2014).
\newblock {Optimal control of transient dynamics in balanced networks supports
  generation of complex movements}.
\newblock {\em Neuron}, 82(6):1394--1406.

\bibitem[Hopfield, 1982]{Hopfield1982}
Hopfield, J.~J. (1982).
\newblock {Neural networks and physical systems with emergent collective
  computational abilities.}
\newblock {\em Proceedings of the National Academy of Sciences of the United
  States of America}, 79(8):2554--2558.

\bibitem[Hornik et~al., 1989]{Hornik1989}
Hornik, K., Stinchcombe, M., and White, H. (1989).
\newblock {Multilayer feedforward networks are universal approximators}.
\newblock {\em Neural Networks}, 2(5):359--366.

\bibitem[Jaeger, 2001]{Jaeger}
Jaeger, H. (2001).
\newblock {The “ echo state ” approach to analysing and training recurrent
  neural networks – with an Erratum note 1}.
\newblock {\em GMD Report}, (148):1--47.

\bibitem[Kadmon and Sompolinsky, 2015]{Kadmon2015}
Kadmon, J. and Sompolinsky, H. (2015).
\newblock Transition to chaos in random neuronal networks.
\newblock {\em Physical Review X}, 5(4):041030.

\bibitem[Leshno et~al., 1993]{Leshno1993}
Leshno, M., Lin, V.~Y., Pinkus, A., and Schocken, S. (1993).
\newblock {Multilayer feedforward networks with a nonpolynomial activation
  function can approximate any function}.
\newblock {\em Neural Networks}, 6(6):861--867.

\bibitem[Logiaco et~al., 2019]{Logiaco2019}
Logiaco, L., Abbott, L., and Escola, S. (2019).
\newblock {A model of flexible motor sequencing through thalamic control of
  cortical dynamics}.
\newblock {\em bioRxiv}, page 2019.12.17.880153.

\bibitem[Maass et~al., 2007]{Maass2007}
Maass, W., Joshi, P., and Sontag, E.~D. (2007).
\newblock {Computational aspects of feedback in neural circuits}.
\newblock {\em PLoS Comput Biol}, 3(1):165.

\bibitem[Maass et~al., 2002]{Maass2002}
Maass, W., Natschl{\"{a}}ger, T., and Markram, H. (2002).
\newblock {Real-time computing without stable states: A new framework for
  neural computation based on perturbations}.
\newblock {\em Neural Computation}, 14(11):2531--2560.

\bibitem[Machens et~al., 2010]{Machens2010}
Machens, C.~K., Romo, R., and Brody, C.~D. (2010).
\newblock {Functional, but not anatomical, separation of "what" and "when" in
  prefrontal cortex}.
\newblock {\em Journal of Neuroscience}, 30(1):350--360.

\bibitem[Mante et~al., 2013]{Mante2013}
Mante, V., Sussillo, D., Shenoy, K.~V., and Newsome, W.~T. (2013).
\newblock {Context-dependent computation by recurrent dynamics in prefrontal
  cortex}.
\newblock {\em Nature}, 503(7474):78--84.

\bibitem[Mastrogiuseppe and Ostojic, 2018]{Mastrogiuseppe2018}
Mastrogiuseppe, F. and Ostojic, S. (2018).
\newblock {Linking Connectivity, Dynamics, and Computations in Low-Rank
  Recurrent Neural Networks}.
\newblock {\em Neuron}, 99(3):609--623.e29.

\bibitem[Mastrogiuseppe and Ostojic, 2019]{Mastrogiuseppe2019}
Mastrogiuseppe, F. and Ostojic, S. (2019).
\newblock {A geometrical analysis of global stability in trained feedback
  networks}.
\newblock {\em Neural Computation}, 31(6):1139--1182.

\bibitem[Nakatsukasa, 2019]{Nakatsukasa2019}
Nakatsukasa, Y. (2019).
\newblock {The low-rank eigenvalue problem}.
\newblock {\em arXiv}, page 1905.11490.

\bibitem[Park and Sandberg, 1991]{Park1991}
Park, J. and Sandberg, I.~W. (1991).
\newblock {Universal Approximation Using Radial-Basis-Function Networks}.
\newblock {\em Neural Computation}, 3(2):246--257.

\bibitem[Pollock and Jazayeri, 2019]{Pollock2019}
Pollock, E. and Jazayeri, M. (2019).
\newblock {Engineering recurrent neural networks from task-relevant manifolds
  and dynamics}.
\newblock {\em bioRxiv}, page 2019.12.19.883207.

\bibitem[Rabinovich et~al., 2008]{Rabinovich2008}
Rabinovich, M., Huerta, R., and Laurent, G. (2008).
\newblock {Neuroscience: Transient dynamics for neural processing}.
\newblock {\em Science}, 321(5885):48--50.

\bibitem[Rajan et~al., 2016]{Rajan2016}
Rajan, K., Harvey, C.~D., and Tank, D.~W. (2016).
\newblock {Recurrent Network Models of Sequence Generation and Memory}.
\newblock {\em Neuron}, 90(1):128--142.

\bibitem[Remington et~al., 2018a]{Remington2018a}
Remington, E.~D., Egger, S.~W., Narain, D., Wang, J., and Jazayeri, M. (2018a).
\newblock {A Dynamical Systems Perspective on Flexible Motor Timing}.
\newblock {\em Trends in Cognitive Sciences}, 22(10):938--952.

\bibitem[Remington et~al., 2018b]{Remington2018b}
Remington, E.~D., Narain, D., Hosseini, E.~A., and Jazayeri, M. (2018b).
\newblock {Flexible Sensorimotor Computations through Rapid Reconfiguration of
  Cortical Dynamics}.
\newblock {\em Neuron}, 98(5):1005--1019.e5.

\bibitem[Rigotti et~al., 2013]{Rigotti2013}
Rigotti, M., Barak, O., Warden, M.~R., Wang, X.~J., Daw, N.~D., Miller, E.~K.,
  and Fusi, S. (2013).
\newblock {The importance of mixed selectivity in complex cognitive tasks}.
\newblock {\em Nature}, 497(7451):585--590.

\bibitem[Rivkind and Barak, 2017]{Rivkind2017}
Rivkind, A. and Barak, O. (2017).
\newblock {Local Dynamics in Trained Recurrent Neural Networks}.
\newblock {\em Physical Review Letters}, 118(25):258101.

\bibitem[Saxena and Cunningham, 2019]{Saxena2019}
Saxena, S. and Cunningham, J.~P. (2019).
\newblock {Towards the neural population doctrine}.
\newblock {\em Current Opinion in Neurobiology}, 55:103--111.

\bibitem[Schuessler et~al., 2020a]{Schuessler2020}
Schuessler, F., Dubreuil, A., Mastrogiuseppe, F., Ostojic, S., and Barak, O.
  (2020a).
\newblock {Dynamics of random recurrent networks with correlated low-rank
  structure}.
\newblock {\em Physical Review Research}, 2(1):013111.

\bibitem[Schuessler et~al., 2020b]{Schuessler2020a}
Schuessler, F., Mastrogiuseppe, F., Dubreuil, A., Ostojic, S., and Barak, O.
  (2020b).
\newblock {The interplay between randomness and structure during learning in
  RNNs}.
\newblock {\em arXiv}, page 2006.11036.

\bibitem[Sohn et~al., 2019]{Sohn2019}
Sohn, H., Narain, D., Meirhaeghe, N., and Jazayeri, M. (2019).
\newblock {Bayesian Computation through Cortical Latent Dynamics}.
\newblock {\em Neuron}, 103(5):934--947.e5.

\bibitem[Sussillo and Abbott, 2009]{Sussillo2009}
Sussillo, D. and Abbott, L. (2009).
\newblock {Generating Coherent Patterns of Activity from Chaotic Neural
  Networks}.
\newblock {\em Neuron}, 63(4):544--557.

\bibitem[Sussillo et~al., 2015]{Sussillo2015}
Sussillo, D., Churchland, M.~M., Kaufman, M.~T., and Shenoy, K.~V. (2015).
\newblock {A neural network that finds a naturalistic solution for the
  production of muscle activity}.
\newblock {\em Nature Neuroscience}, 18(7):1025--1033.

\bibitem[Vyas et~al., 2020]{Vyas2020}
Vyas, S., Golub, M.~D., Sussillo, D., and Shenoy, K.~V. (2020).
\newblock {Computation Through Neural Population Dynamics}.
\newblock {\em Annual Review of Neuroscience}, 43(1):249--275.

\bibitem[Wang et~al., 2018]{Wang2018}
Wang, J., Narain, D., Hosseini, E.~A., and Jazayeri, M. (2018).
\newblock {Flexible timing by temporal scaling of cortical responses}.
\newblock {\em Nature Neuroscience}, 21(1):102--112.

\bibitem[Yang et~al., 2019]{Yang2019}
Yang, G.~R., Joglekar, M.~R., Song, H.~F., Newsome, W.~T., and Wang, X.-J.
  (2019).
\newblock {Task representations in neural networks trained to perform many
  cognitive tasks}.
\newblock {\em Nature Neuroscience}, 22(2):297--306.

\end{thebibliography}
\newpage

\begin{appendices}

\section{Dynamics in multi-population networks}\label{app:1}

    In this appendix, we derive the equation for the dynamics of a multi-population low-rank network, Eq.~\eqref{dyn_pop_general_cont}. We consider a low-rank network that consists of $P$ populations, where each population is defined by different statistics of the probability distribution $P_{\left(p\right)}\left(\underline{m}, \underline{n}, \underline{I}\right)$. We assume that the external input is constant in time and uncorrelated with the left connectivity patterns at the level of each population. Each neuron in the network is assigned to a population according to the probability $\alpha_p$. In the following, we set the statistics of each population to be drawn from a multivariate Gaussian with mean vector $\boldsymbol{a^{\left(p\right)}}$, as defined in Eq.~\eqref{def_mu}, and covariance matrix $\Sigma^{\left(p\right)}$ (Eq.~\ref{corr_gen}).

    The recurrent dynamics in a low-rank network are determined by Eq.~\eqref{rec_dyn}: it consists of a sum over the $N$ units in the network. In the limit of large networks with defined statistics, by means of the law of large numbers, this sum over $N$ i.i.d. elements corresponds to the empirical average over the distribution of its elements. Therefore, we can replace the sum over network units for $i=1, \dots, N$ of loadings $\left\{n_i^{\left(r\right)}\right\}, \left\{m_i^{\left(r\right)}\right\}$ and $I_i$, by an integral over their  probability distribution $P\left(\underline{m}, \underline{n}, I\right)$. Using this probability distribution, the recurrent dynamics in Eq.~\eqref{rec_dyn} can be expressed as
    
    \begin{align} \label{rec_int}
        \kappa_r^{rec} &=  \sum_{p=1}^P \alpha_p \int d\underline{m}\, d\underline{n}\, dI P_{\left(p\right)}\left(\underline{m}, \underline{n}, I  \right) n_r^{\left(p\right)} \phi\left(I^{\left(p\right)} \kappa_I + \sum_{l=1}^R m_l^{\left(p\right)} \kappa_l \right).
    \end{align}

    \noindent Note that we refer to the input loadings $I$ as a single Gaussian variable, instead of a set of Gaussian variables $\underline{I}$, because there can only be one effective input pattern when the input is constant in time. We then separate the contribution of the mean $a_{n_r}$ and the fluctuations of $n_r$ around its mean into two different terms:
\begin{align} \label{rec_int_1}
        \kappa_r^{rec} = & \sum_{p=1}^{P} \alpha_p \int dI\, d\underline{m}\,  P_{\left(p\right)}\left(\underline{m}, I\right) a_{n_r}^{\left(p\right)} \phi \left(I^{\left(p\right)}\,\kappa_I  + \sum_{l=1}^R m_l^{\left(p\right)} \kappa_l\right) \nonumber \\
         + &\sum_{p=1}^{P} \alpha_p \int d n_r\,dI\, d\underline{m}\,  P_{\left(p\right)}\left(\underline{m}, n_r, I\right) \left(n_r^{\left(p\right)} - a_{n_r}^{\left(p\right)} \right) \phi \left(I^{\left(p\right)}\,\kappa_I  + \sum_{l=1}^R m_l^{\left(p\right)} \kappa_l\right).
    \end{align}

    This derivation is implicitly conditioned on the values of the collective variables $\kappa_r$. Under such conditioning, the argument inside the function $\phi$ in Eq.\eqref{rec_int_1} is itself a Gaussian variable. Using Stein's lemma in the second term, and expressing the argument of the transfer function as a single Gaussian variable, we can express the dynamics as
    \begin{align} \label{rec_int2}
        \kappa_r^{rec} &= \sum_{p=1}^P \alpha_p  a_{n_r}^{\left(p\right)}\int \mathcal{D}x  \, \phi \left(a_I^{\left(p\right)} \kappa_I+ \sum_{s=1}^R a_{m_s}^{\left(p\right)} \kappa_s + x \sqrt{\sigma_{I^2}^{\left(p\right)}\kappa_I^2 +\sum_{s^\prime=1}^R  \sigma_{m_{s^\prime}^2}^{\left(p\right)} \kappa_{s^\prime}^2}  \right) \nonumber \\
        &+\sum_{p=1}^P \alpha_p \left( \sigma_{n_r I}^{\left(p\right)} \kappa_I + \sum_{s=1}^R \sigma_{n_r m_s}^{\left(p\right)} \kappa_s\right) \int \mathcal{D}x \,\phi^\prime \left(a_I^{\left(p\right)} \kappa_I+ \sum_{s=1}^R a_{m_s}^{\left(p\right)} \kappa_s + x \sqrt{\sigma_{I^2}^{\left(p\right)}\kappa_I^2 +\sum_{s^\prime=1}^R  \sigma_{m_{s^\prime}^2}^{\left(p\right)} \kappa_{s^\prime}^2}\right)
    \end{align}
    \noindent where $\mathcal{D}x = dx \left(2\pi\right)^{-\frac{1}{2}} e^{-\frac{x^2}{2}}$. Finally, using the Gaussian integral notation in Eq.~\eqref{notation}, we retrieve Eq.~\eqref{rec_dyn_pop}.

\section{Universal approximation of low-dimensional dynamics}\label{app:2}
         The universal approximation theorem for artificial neural networks \citep{ Hornik1989, Funahashi1989, Cybenko1989} states that any piecewise-continuous bounded function $G\left(\boldsymbol{x}\right)$, where $\boldsymbol{x}$ is a $d$-dimensional vector, can be approximated to arbitrary precision by a finite linear combination of non-linear units having the same transfer function but different gain and thresholds. More precisely, it is possible to build an approximation $\hat{G}\left(\boldsymbol{x}\right)$ of $G\left(\boldsymbol{x}\right)$  
        \begin{equation}\label{approx}
        \hat{G}\left(\boldsymbol{x}\right) = \sum_{i=1}^N \boldsymbol{v}_i\phi \left(\boldsymbol{w}_i^T \boldsymbol{x} + b_i\right),
        \end{equation}
          with finite integer $N$, and real values for $\boldsymbol{v}_i \in \mathcal{R}^{d^\prime}$, $\boldsymbol{w}_i \in \mathcal{R}^d$ and $b_i \in \mathcal{R}$, so that $\left\lvert G\left(x\right)-\hat{G}\left(x\right)\right\rvert<\epsilon$, for any $\epsilon >0$, given mild assumptions on the non-linear activation function $\phi\left(x\right)$. Historically, the universality property has been shown using a wide range of transfer functions, using initially squashing or sigmoidal functions \citep{Funahashi1989, Hornik1989, Cybenko1989}, inclusing Heaviside functions, sinusoidal functions \citep{Gallant1988} and radial basis functions \citep{Park1991}. Years later, it was shown that the necessary and sufficient condition on the transfer function for the universal approximation property is that $\phi\left(x\right)$  be a non-polynomial function \citep{Leshno1993}. 
          
         There is a direct mapping between the second term of Eq.~\eqref{approx} and the recurrent dynamics of a low-rank RNN. The recurrent dynamics in Eq.~\eqref{rec_dyn} can be directly mapped to Eq.~\eqref{approx}: the variables $\frac{1}{N}\boldsymbol{n}_i$ correspond to $\boldsymbol{v}_i$, $\boldsymbol{m}_i$ to $\boldsymbol{w}_i$, and $\kappa_I I_i$ to  $b_i$. Thitatis implies that the recurrent dynamics can approximate any flow function within a finite domain. The parameters $\frac{1}{N}\boldsymbol{n}_i$, $\boldsymbol{m}_i$, and $\kappa_I I_i$ can be adjusted independently. In particular, $\kappa_I I_i$ is independent of all the other collective variables $\kappa_r$, although the opposite is not true: the collective variables $\kappa_r$ obviously depend on the external tonic input $\kappa_I I_i$. 
         
         The dynamics of low-rank networks with multiple Gaussian populations can also be mapped to the universal approximation theorem. The mean term contribution to the dynamics in Eq.~\eqref{dyn_pop_general_cont} reads 
          \begin{align} \label{dyn_pop_general2}
         \sum_{p=1}^P \alpha_p  \boldsymbol{a_{n}}^{\left(p\right)} \left\langle \phi \left(\boldsymbol{a_m}^T \boldsymbol{\kappa} + a_I^{\left(p\right)}, \sigma_{I^2}^{\left(p\right)} + \boldsymbol{\kappa}^T \sigma_{m^2}^{\left(p\right)} \boldsymbol{\kappa} \right) \right\rangle,
    \end{align}
    \noindent so that $\alpha_p \boldsymbol{a_{n}}^{\left(p\right)}$ maps to $\boldsymbol{v_i}$, $\boldsymbol{a_{m}}^{\left(p\right)}$ maps to $\boldsymbol{w}_i$ and $a_I^{\left(p\right)}$ is mapped to the bias term $b_i$. The transfer function is however different. In Eq.~\eqref{approx}, the non-linear function used is $\phi\left(x\right)$, while in Eq.~\eqref{dyn_pop_general2}, the non-linear function used is $\left\langle \phi\left(x, \Delta\left(x\right)\right)\right\rangle$. Both functions are non-linear and non-polynomial, so that the theorem applies in each case. The contribution given by the disorder in the population loadings, $\boldsymbol{\sigma}_{m^2}^{\left(p\right)}$ and $\sigma_{I^2}^{\left(p\right)}$ are not required for the universal approximation. However, quadratic terms like the one introduced by the variance of loadings improve the approximation in terms of expressibility and efficiency \citep{Fan2020}. Overall, this means that a low-rank network with a finite number of populations can approximate any dynamical system within a bounded domain.
    
\section{Linear stability matrix at fixed points in networks with single population}\label{app:3}
    The linear dynamics of small perturbations around the fixed point $\boldsymbol{\kappa_0}$ (defined in Eqs.~\ref{dyn_vec_fp}) read

    \begin{equation}\label{linear}
        \tau \frac{d\boldsymbol{\kappa}}{dt} = - \boldsymbol{\kappa} + \left[\nabla \left( \left\langle \phi^\prime \left(0, \boldsymbol{\kappa}^T \boldsymbol{\kappa}\right) \right\rangle\boldsymbol{\sigma_{mn}} \boldsymbol{\kappa}\right)\right]_{\boldsymbol{\kappa}=\boldsymbol{\kappa_0}} \boldsymbol{\kappa},
    \end{equation}
    \noindent where $\nabla$ is the vector differential operator. We apply the property $\nabla\left(f\left(\boldsymbol{\kappa}\right) A \boldsymbol{\kappa})\right) = f\left(\boldsymbol{\kappa}\right) A + A\boldsymbol{\kappa} \left(\nabla f\left(\boldsymbol{\kappa}\right)\right)^T$, based on the chain rule, where $A$ is an $R\times R$ matrix, to obtain:
    
    \begin{equation}\label{int_0}
        \tau \frac{d\boldsymbol{\kappa}}{dt} = - \boldsymbol{\kappa} + \left[\left\langle \phi^\prime \left(0, \boldsymbol{\kappa}^T \boldsymbol{\kappa}\right) \right\rangle\boldsymbol{\sigma_{mn}} +   \boldsymbol{\sigma_{mn}} \boldsymbol{\kappa}  \left\langle \nabla \phi^\prime \left(0, \boldsymbol{\kappa}^T \boldsymbol{\kappa}\right) \right\rangle^T\right]_{\boldsymbol{\kappa}=\boldsymbol{\kappa_0}} \boldsymbol{\kappa}.
    \end{equation}
    
    \noindent We then calculate the gradient of the gain factor. To do so, we first write explicitly the Gaussian integral
    \begin{equation}\label{int_1}
        \left\langle \nabla \phi^\prime \left(0, \boldsymbol{\kappa}^T \boldsymbol{\kappa}\right) \right\rangle = \int \mathcal{D}x  \nabla \phi^{\prime} \left( \sqrt{\boldsymbol{\kappa}^T \boldsymbol{\kappa}} x \right), 
    \end{equation}
    \noindent where $\mathcal{D}x$ is the differential element of a normally distributed variable. Applying the chain rule
    \begin{align}
        \left\langle \nabla \phi^\prime \left(0, \boldsymbol{\kappa}^T \boldsymbol{\kappa}\right) \right\rangle &= \int \mathcal{D}x   \phi^{\prime\prime} \left( \sqrt{\boldsymbol{\kappa}^T \boldsymbol{\kappa}} x \right) \nabla\left(x \sqrt{\boldsymbol{\kappa}^T \boldsymbol{\kappa}} \right)= \int \mathcal{D}x   \phi^{\prime\prime} \left( \sqrt{\boldsymbol{\kappa}^T \boldsymbol{\kappa}} x \right) x \frac{\boldsymbol{\kappa}}{\sqrt{\boldsymbol{\kappa}^T \boldsymbol{\kappa} }}.
    \end{align}
    \noindent Using Stein's lemma, the gradient of the gain factor reads:
    \begin{align}
        \left\langle \nabla \phi^\prime \left(0, \boldsymbol{\kappa}^T \boldsymbol{\kappa}\right) \right\rangle 
        &= \int \mathcal{D}x   \phi^{\prime\prime} \left( \sqrt{\boldsymbol{\kappa}^T \boldsymbol{\kappa}} x \right) \frac{\boldsymbol{\kappa}}{\sqrt{\boldsymbol{\kappa}^T \boldsymbol{\kappa}}}=\int \mathcal{D}x   \phi^{\prime\prime\prime} \left( \sqrt{\boldsymbol{\kappa}^T \boldsymbol{\kappa}} x \right) \boldsymbol{\kappa}= \left\langle \phi^{\prime \prime \prime} \left(0, \boldsymbol{\kappa}^T \boldsymbol{\kappa}\right)\right\rangle \boldsymbol{\kappa}.\label{sol_third}
    \end{align}
    
    \noindent Finally, introducing Eq.~\eqref{sol_third} into Eq.~\eqref{int_0}, and using the fact that $\boldsymbol{\sigma_{mn}} \boldsymbol{\kappa_0} = \lambda_r \boldsymbol{\kappa_0}$, the dynamics of small perturbation around the fixed point read
        \begin{equation}
        \tau \frac{d\boldsymbol{\kappa}}{dt} =  \left[ -\boldsymbol{I} + \left\langle \phi^\prime \left(0, \boldsymbol{\kappa_0}^T \boldsymbol{\kappa_0}\right) \right\rangle\boldsymbol{\sigma_{mn}} + \left\langle \phi^{\prime \prime \prime} \left(0, \boldsymbol{\kappa_0}^T \boldsymbol{\kappa_0}\right)\right\rangle  \boldsymbol{\sigma_{mn}} \boldsymbol{\kappa_0}\boldsymbol{\kappa_0}^T  \right] \boldsymbol{\kappa},
    \end{equation}

    \noindent which leads to the linear stability matrix given by Eq.~\eqref{lin_stab}.
    
    It is important to analyze the behavior of the function $\left\langle \phi^{\prime \prime \prime} \left(0, \Delta\right)\right\rangle$ to assess the stability. In the limit $\Delta =0$, the Gaussian integral reduces to the evaluation of the function at zero. For a transfer function $\phi\left(x\right)=\tanh\left(x\right)$ we obtain:
    \begin{equation}
        \lim_{\Delta\to 0}\left\langle \phi^{\prime \prime \prime} \left(0, \Delta\right)\right\rangle = \phi^{\prime \prime \prime}\left(0\right) = -2.
    \end{equation}
    In the limit of infinite $\Delta$, the Gaussian integral can be expressed as :
    \begin{equation}
        \lim_{\Delta\to \infty}\left\langle \phi^{\prime \prime \prime} \left(0, \Delta\right)\right\rangle = \int_{-\infty}^{+\infty} dx\,\phi^{\prime \prime \prime}\left(x\right) = 0.
    \end{equation}
    
    Furthermore, it can be shown analytically that $\left\langle \phi^{\prime \prime \prime} \left(0, \Delta\right) \right\rangle$ is never zero for any finite value $\Delta$, by studying the minima of its primitive function. The primitive function of $\left\langle\phi^{\prime \prime \prime} \left(0, \Delta\right) \right\rangle$ is proportional to $\left\langle\phi^{\prime} \left(0, \Delta\right) \right\rangle$. The primitive function has no local minima, because it is a function bounded between 0 and 1 that is a monotonically decreasing function of $\Delta$. We can show that the function is monotonically decreasing because for sigmoidal transfer functions $\phi^\prime\left(\sqrt{\Delta_1}x\right) < \phi^\prime\left(\sqrt{\Delta_2}x\right)$ if and only if $\Delta_1 < \Delta_2$. Thus, this property is still conserved when calculating the Gaussian average: $\left\langle \phi^\prime\left(0, \Delta_1\right) \right\rangle< \left\langle \phi^\prime\left(0, \Delta_2\right) \right\rangle$ if and only if $\Delta_1 < \Delta_2$.
    
    Putting these analyses together, we conclude that the function $\left\langle \phi^{\prime \prime \prime} \left(0, \Delta\right) \right\rangle$ is $-2$ for $\Delta=0$, is always smaller than zero, and tends asymptotically to this upper bound as $\Delta$ approaches infinity. This result is used to study whether the linearized dynamics around fixed points in low-rank networks with a single Gaussian population are stable (Eq.~\ref{stab}).
    
    
    

\section{Stability analysis of rank-two networks with non-normal covariance}\label{app:4}

In Section 4, we analyzed the dynamics generated by a rank-two network with one single Gaussian population when the covariance matrix $\boldsymbol{\sigma_{mn}}$ is \textit{normal}, i.e., the eigenvectors are mutually orthogonal to each other. We extend here the analysis to non-normal matrices, and show that the main features of the dynamics are conserved when the correlation matrix is non-normal.

We first studied the case of normal matrices with real eigenvalues (Fig.~\ref{fig:T2} A-D, and Eqs.~\eqref{dyn_vec_fp} -~\eqref{stab}). The analysis showed that each real eigenvector $\mathbf{u_r}$ of the covariance matrix $\boldsymbol{\sigma_{mn}}$ leads to a pair of fixed points, when the associated eigenvalue $\lambda_r$ is larger than one. 

The linear dynamics around fixed points is given by the Jacobian $S_r$ in Eq.~\eqref{lin_stab}, where $r=1,\dots,R$. The eigenvectors of $S_r$ coincide with the eigenvectors of $\boldsymbol{\sigma_{mn}}$ if the covariance matrix has real eigenvectors mutually orthogonal to each other. Using this property, we can then determine the eigenvalues of the linearized dynamics around each fixed point in Eq.~\eqref{stab}. As a conclusion, this analysis showed that all fixed points are saddle points, except for the two fixed points in the direction of the eigenvector with largest associated eigenvalue, where the fixed points are stable. 

We now extend this analysis to networks where the covariance matrix $\boldsymbol{\sigma_{mn}}$ is a \textit{non-normal}  matrix with real eigenvalues, in other words,  a matrix  with non-zero correlations between eigenvectors. Each real eigenvector $\mathbf{u_r}$ with eigenvalue $\lambda_r>1$ generates two fixed points in the direction it spans, exactly the same way as in matrices with orthogonal eigenvectors. The fixed points are also located at the same radial distance along each eigenvector (Eq.~\ref{scal_fp}), compared to the case of normal matrices. The Jacobian matrix at the fixed point is still given by Eq.~\eqref{lin_stab}:

     \begin{equation}\label{lin_stab_ap1}
         S_r = -\boldsymbol{I} +   \frac{1}{\lambda_r} \boldsymbol{\sigma_{mn}} + \left\langle \phi^{\prime\prime\prime} \left(0, \rho_r^2\right) \right\rangle \lambda_r \rho_r^2 \boldsymbol{u_r}\boldsymbol{u_r}^T.
     \end{equation}
     
\noindent When the covariance matrix $\boldsymbol{\sigma_{mn}}$ is a non-normal matrix , the eigenvectors of $S_r$ are not equal to the eigenvectors of the covariance. Finding an analytical expression for the eigenvectors of $S_r$ is in general a challenging problem. However, we can still calculate the eigenvalues of $S_r$ and show that they remain unchanged in networks with normal covariances (Eq.~\ref{stab}). 

Let's consider the plane spanned by the eigenvector $\mathbf{u_r}$ and any other eigenvector of the covariance matrix $\mathbf{u_{r^\prime}}$. We consider only the linearized dynamics around the fixed point along this plane. We project $S_r$ onto this plane, and indicate it with the notation $\left[S_r\right]$, which is a $2\times2$ block sub-matrix of the full Jacobian $S_r$.

     \begin{equation}\label{lin_stab_ap2}
         \left[S_r\right] = -\left[\boldsymbol{I}\right] +   \frac{1}{\lambda_r} \left[\boldsymbol{\sigma_{mn}}\right] + \left\langle \phi^{\prime\prime\prime} \left(0, \rho_r^2\right) \right\rangle \lambda_r \rho_r^2 \boldsymbol{u_r}\boldsymbol{u_r}^T.
     \end{equation}
The vector $\mathbf{u_r}$ is still an eigenvector of $\left[S_r\right]$. Its associated eigenvalue is

\begin{equation}\label{eq_stab1}
    \gamma_r = \left\langle \phi^{\prime\prime\prime} \left(0, \rho_r^2\right) \right\rangle \lambda_r \rho_r^2
\end{equation}
 which is a negative number. This implies that all fixed points are stable in the radial direction. Although calculating the second eigenvector is not obvious, we can calculate the second eigenvalue of $\left[S_r\right]$ by calculating its trace and subtracting $\gamma_r$:

\begin{equation}
    \gamma_{r^\prime} = \text{tr}\left(\left[S_r\right]\right) - \gamma_r.
\end{equation}

\noindent Using the linearity of the trace operator, the expression simplifies to
\begin{equation}\label{eq_stab2}
    \gamma_{r^\prime} = -1 + \frac{\lambda_{r^\prime}}{\lambda_r} .
\end{equation}

\noindent Therefore, the eigenvalues of the full Jacobian $S_r$, given by Eqs.~\eqref{eq_stab1} and ~\eqref{eq_stab2}, coincide with the eigenvalues of the Jacobian in the case of normal matrices (Eq.~\ref{stab}). The fixed points along the eigenvector direction $\mathbf{u_r}$ are stable if $\lambda_r$ is the largest eigenvalue of $\mathbf{\sigma_{mn}}$. Otherwise, the fixed point is a saddle point.

\begin{figure}[h]
     \centering
     \begin{subfigure}[b]{\textwidth}
        \includegraphics[width=\linewidth]{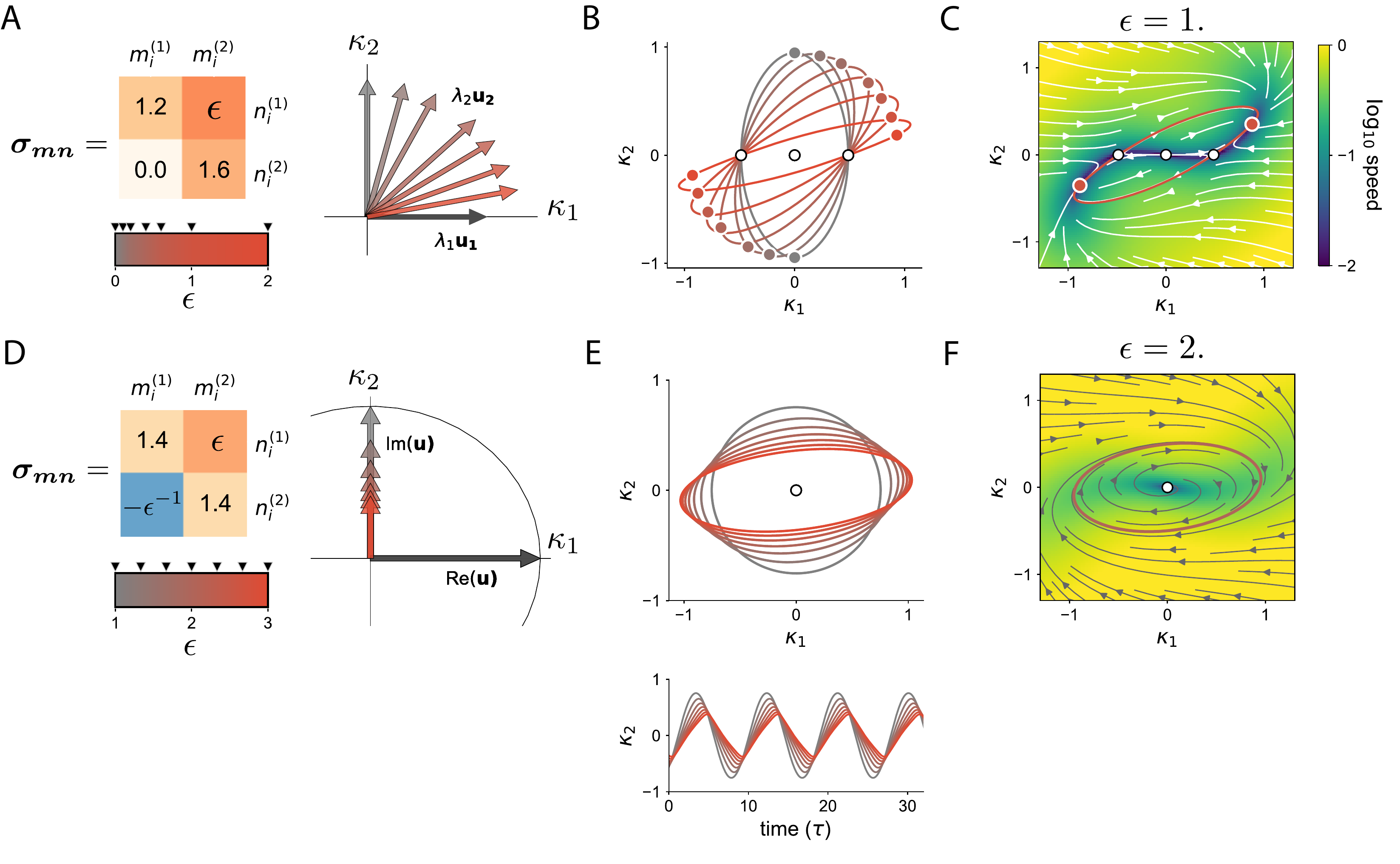}
     \end{subfigure}
	   	\caption{\textbf{Dynamics in rank-two networks with non-normal covariances $\boldsymbol{\sigma_{mn}}$}. \textbf{A} Covariance matrix $\sigma_{mn}$ with eigenvalues $\lambda_1 = 1.2$ and $\lambda_2 = 1.6$. The free parameter $\epsilon$ controls the angle between the eigenvectors, as shown in the right. \textbf{B} Stable fixed points (colored dots) and saddle points (white dots) generated by the non-normal covariance matrix for different values of $\epsilon$. The colored lines indicate the trajectories between saddle points and stable fixed points. As $\epsilon$ increases, the location of the stable fixed points move closer to the horizontal line, while keeping the same radial distance. The stability of the fixed points does not change with $\epsilon$. \textbf{C} Example of the full dynamics in collective fixed space for a fixed value of $\epsilon=1$. \textbf{D} Covariance matrix $\sigma_{mn}$ with eigenvalues $\lambda = 1.4 \pm 1$. The free parameter $\epsilon>1$ controls the relative norm of the imaginary part of the eigenvector $\mathbf{u}$, without modifying the fixed points. \textbf{E} Top. Limit cycles that emerge in the dynamics for different values of $\epsilon$. When the covariance matrix is normal ($\epsilon=1$, grey line), the limit cycle has a circular shape. As $\epsilon$ increases, the limit cycle loses the circular symmetry and resembles an ellipse. Bottom. Projection of the collective variable $\kappa_2$ as a function of time. As $\epsilon$ increases, the activity loses its sinusoidal shape, while keeping the same frequency. \textbf{F} Example of the dynamical landscape in collective space for a fixed value of $\epsilon=2$. } \label{fig:SFig1}
    \end{figure}

Figure~\ref{fig:SFig1} A-C shows how the dynamics vary when the correlation matrix $\boldsymbol{\sigma_{mn}}$ is non-normal, but the eigenvalues remain constant. The parameter $\epsilon$  determines the degree of correlation between eigenvectors: as $\epsilon$ increases, the two eigenvectors become more strongly correlated (Fig.~\ref{fig:SFig1} A). This correlation moves the direction of the associated fixed point, while keeping the radial distance constant (Fig.~\ref{fig:SFig1} B). The eigenvalues of the linear dynamics around the fixed points do not vary. Fig.~\ref{fig:SFig1} C shows an example of the dynamical landscape in collective space. It is interesting to note that the trajectories that go from the saddle points towards the stable fixed points resemble  ellipses, similarly to the case of normal $\boldsymbol{\sigma_{mn}}$ (Fig.~\ref{fig:T2} A-D). However, the stable fixed points are no longer located in the furthest point of this curve from the origin.

In summary, we showed that the number of fixed points and the stability in a low-rank network with one single population depends only on the eigenvalues of $\mathbf{\sigma_{mn}}$, and not on its eigenvectors. When the eigenvectors are all mutually orthogonal, we can calculate analytically the eigenvalues and eigenvectors of the linearized dynamics around each fixed point. When the eigenvectors are not all mutually orthogonal, the eigenvectors of the linearized dynamics change, but not the eigenvalues. This suggests that increasing the correlation between eigenvectors introduces a continuous deformation of the collective space while all the temporal quantities are preserved.

We observed numerically that the previous result holds as well for covariance matrices $\boldsymbol{\sigma_{mn}}$ with complex eigenvalues. Each pair of complex conjugate eigenvalues $\lambda\pm i \omega$ with eigenvector $\mathbf{v_1} + i \mathbf{v_2}$ generates a limit cycle on the plane spanned by $\mathbf{v_1}$\textendash$\mathbf{v_2}$ if $\lambda>1$. If $\boldsymbol{\sigma_{mn}}$ is a normal matrix, such that vectors $\mathbf{v_1}$ and $\mathbf{v_2}$ have the same norm and are orthogonal to each other, the limit cycle is a circle. In that case, we showed in Section 4 that the frequency of the limit cycle is given by $\frac{\omega}{\lambda}$. 

Adding correlations between the real and imaginary part of the eigenvectors $\mathbf{v_1}$ and $\mathbf{v_2}$, or changing their relative norm, will change the shape of the limit cycle to a curve resembling an ellipse. Fig.~\ref{fig:SFig1} D-F shows an example of a covariance matrix $\boldsymbol{\sigma_{mn}}$ where the non-normality is controlled by parameter $\epsilon$, without affecting the eigenvalues. As $\epsilon$ increases, the relative norm of the imaginary and real part of the eigenvector $\mathbf{u}$ changes (Fig.~\ref{fig:SFig1} D). This correlation changes the shape of the limit cycle, from a circle when $\epsilon$ is one (normal case, Fig.~\ref{fig:SFig1} E grey line) to a closed curve resembling an ellipse as $\epsilon$ increases (Fig.~\ref{fig:SFig1} E-F grey line). The frequency of oscillation along the limit cycle however stays invariant as $\epsilon$ is varied (Fig.~\ref{fig:SFig1} E bottom).

To sum up, also in the case of complex eigenvalues, the correlations between eigenvectors deforms the collective space, introducing a continuous mapping between the circular limit cycle, and the new limit cycle, that resembles an ellipse. However, the temporal features of the dynamical landscape, such as the frequency of the limit cycle, remain constant. 

\begin{figure}[h]
     \centering
     \begin{subfigure}[b]{\textwidth}
        \includegraphics[width=\linewidth]{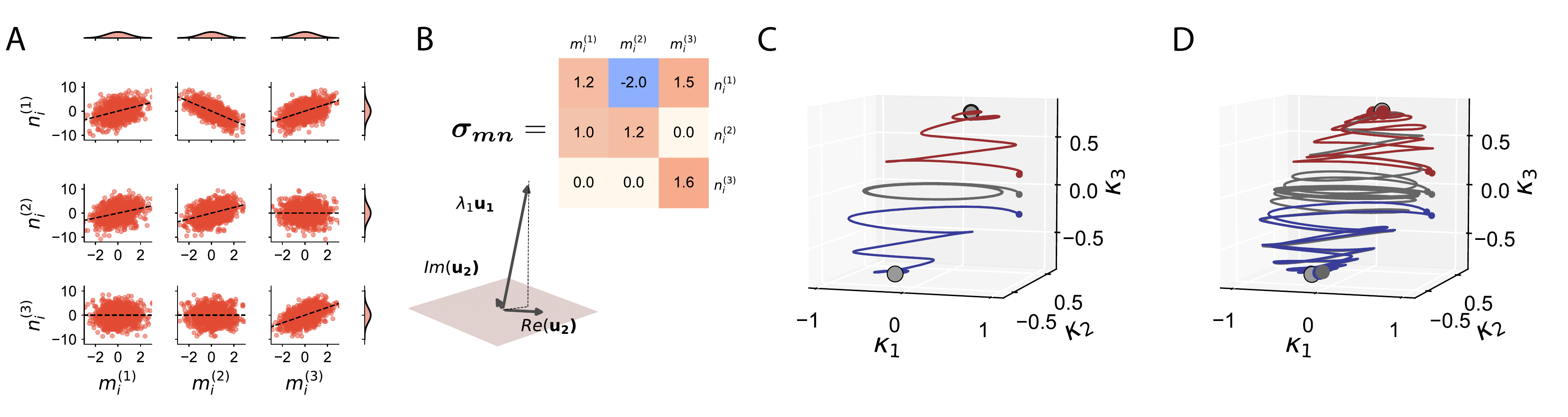}
     \end{subfigure}
	\caption{\textbf{Dynamics in a rank-three network with a single Gaussian population - connectivity matrix combining real and  complex eigenvalues}. \textbf{A} Scatter plot between the loadings of connectivity patterns $m_i^{\left(r\right)}$ and $n_i^{\left(r\right)}$. $\sigma=1.6$ and $\sigma_{\omega}=0.8$. \textbf{B} Covariance matrix of the singular vectors (top) and sketch of the  eigenvectors (bottom). The eigenvalues are $\lambda_1 = 1.6$ and $\lambda_{2,3} = 1.2 \pm \sqrt{2}$. The real eigenvector $\mathbf{u_1}$ is not orthogonal to the plane spanned by the real and imaginary part of the complex eigenvectors $\mathbf{u_2}$. The real and imaginary part of the complex eigenvectors span the horizontal plane (shaded in grey) and do not have the same norm.  \textbf{C} Mean-field dynamics (Eq.~\ref{dyn_vec}) for three trajectories starting at different initial conditions. Each color indicates a different trajectory. When the network is initialized in the horizontal plane (grey trajectory), the activity oscillates within the non-circular limit cycle. Otherwise it converges to one of the two stable fixed points, located in the direction of the eigenvector $\bf{u_1}$. \textbf{D} Same trajectories as in \textbf{G}, in finite-size simulations, for three different connectivity matrices. The trajectories always end up in one of the two stable fixed points, even if initialized in the horizontal plane (grey trajectories). Parameters: $N=1000, \sigma_{n_r^2}=9$.} \label{fig:SFig2}  
    \end{figure}

Real and complex eigenvalues are combined in networks with rank three or larger. The same stability rules hold: the dynamic structure (limit cycle or fixed point) generated by the eigenvalue with largest real part is stable, while the other ones are not stable in all directions. Figure \ref{fig:SFig2} A-B shows an example of a rank-three network, whose connectivity matrix has a real eigenvalue $\lambda_1$ and a pair of complex conjugate eigenvalues $\lambda_2$ and $\lambda_3$. The real part of all eigenvalues is larger than one, so that the real eigenvalue leads to a pair of fixed points, and the complex eigenvalues generate a limit cycle. Given that the real eigenvalue $\lambda_1$ is larger than the real part of the other eigenvalues, the fixed points are stable. The limit cycle is marginally stable in the plane spanned by the real and imaginary parts of the complex eigenvector of $\lambda_{2}$, but unstable in any other direction. Therefore, trajectories starting in the plane converge to the limit cycle in the mean-field equation (see grey trajectory in Fig.~\ref{fig:SFig2} C). Small perturbations, such as those introduced by finite-size effects,  make these trajectories deviate from the limit cycle and converge to one of the two stable fixed points (grey trajectories, Fig.~\ref{fig:SFig2} D).
 
\end{appendices}

\end{document}